# Projective Parallel Single-Pixel Imaging: 3D Structured Light Scanning Under Global Illumination

Yuxi Li*, Hongzhi Jiang*, Huijie Zhao § , and Xudong Li

**Abstract**—We present projective parallel single-pixel imaging (pPSI), a 3D photography method that provides a robust and efficient way to analyze the light transport behavior and enables separation of light effect due to global illumination, thereby achieving 3D structured light scanning under global illumination. The light transport behavior is described by the light transport coefficients (LTC), which contain complete information for a projector–camera pair, and is a 4D data set. However, the capture of LTC is generally time consuming. The 4D LTC in pPSI are reduced to projection functions, thereby enabling a highly efficient data capture process. We introduce the local maximum constraint, which provides constraint for the location of candidate correspondence matching points when projections are captured. Local slice extension (LSE) method is introduced to accelerate the capture of projection functions. Optimization is conducted for pPSI under several situations. The number of projection functions required for pPSI is optimized and the influence of capture ratio in LSE on the accuracy of the correspondence matching points is investigated. Discussions and experiments include two typical kinds of global illuminations: inter-reflections and subsurface scattering. The proposed method is validated with several challenging scenarios, and outperforms the state-of-the-art methods.

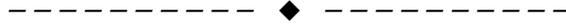

## 1 INTRODUCTION

A common assumption in structure light scanning (SLS) methods, such as fringe projection profilometry (FPP) [1],[2],[3] and Grey coding [4], is that light ray only travels along a direct path when transmitting through a scene. Therefore, SLS methods are susceptible to systematic distortions and random errors when global illumination exists between projector and camera pixels [17],[18],[19],[20],[21]. Global illumination can occur when the investigated objects embody complicated surfaces and materials. For instance, Fig. 1(a) shows that inter-reflections are dominant between highly glossy reflective surfaces. In these surfaces, the light beams received by camera pixels contain not only directly reflected light but also inter-reflected light between surfaces. Fig. 1(b) shows that subsurface scattering effects emerge at translucent surfaces when light penetrates the surface and exits at different positions around the incident point. Analyzing and decomposing the influences caused by global illuminations through modern cameras is a challenging and open problem [15].

Light transport equation describes the complex transport behavior between projector and camera pixels. The path between the projector and camera pixels can be determined by capturing and analyzing the *light transport coefficients* (LTC) [21], which denote the light radiance between every possible projector and combinations of camera positions; this process enables correspondence matching because the direct path can be identified. However, LCTs are a 4D dataset, which parameterizes light rays considering a 2D camera and 2D projector coordinates. Thus, LTC involve enormous data volume and long capturing time. LTC can be visualized by a 2D image with projector resolutions considering a camera pixel. The 2D image with projector resolutions is referred to as *pixel transport image*.

As a step toward a robust and efficient analysis of light transport behavior, we develop projective parallel single-pixel imaging (pPSI) to separate the influences of lights caused by global illumination in 3D scanning. Considering that only the correspondence point is the ultimate goal in the context of 3D reconstruction, LTC contain over-complete information because only the direct correspondence point is extracted and stored for each pixel transport image.

- *: Equal Contribution
- Yuxi Li, Hongzhi Jiang and Xudong Li are with School of Instrumentation and Optoelectronic Engineering, Beihang University (BUAA), Beijing 100191, China.
  E-mail: uniluxli@ qq.com, jhz1862@buaa.edu.cn, xdli@buaa.edu.cn.
- Huijie Zhao is with Institute of Artificial Intelligence, Beihang University (BUAA), Beijing 100191, China. E-mail: hjzhao@buaa.edu.cn.
- § Corresponding author: Huijie Zhao.

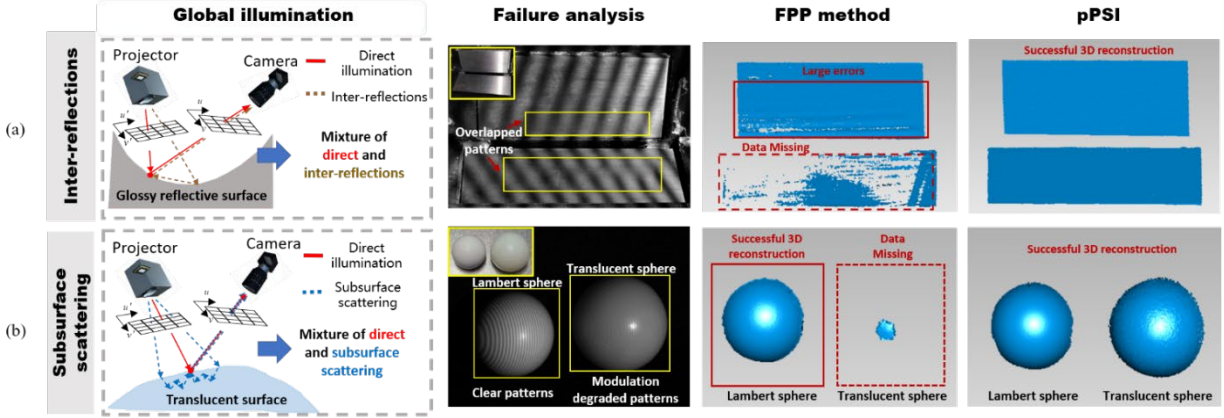

Fig. 1. Global illumination problems discussed in this paper. Typical global illumination effects include (a) inter-reflections and (b) subsurface scattering. Inter-reflections incur overlapped pattern. Subsurface scattering degrades the modulation of the patterns. Global illumination effects cause failure in traditional SLS methods, such as FPP. However, the proposed method (pPSI) can solve these problems robustly and efficiently.

The present paper shows that the correspondence matching can be obtained when the 1D *projection function(s)* of the 2D pixel transport image is captured. Thus, the goal can be achieved by reducing the capture of 4D LTC to projection functions, which largely minimizes the data required to be captured. This process is also robust to correspondence matching for 3D reconstruction in situations dominated by global illuminations.

## 1.1 Contributions

We introduce pPSI, which is a robust, efficient, and comprehensive 3D reconstruction method in the presence of global illumination. Rather than capturing complete 4D LTC, pPSI captures projection functions, thereby enabling highly efficient data capture procedure. The *local maximum constraint* is proven, which states that the corresponding matching point (direct illumination point) on the pixel transport image is retained as the local maximum on the projection function(s), if the projection line through the local maximum does not pass through the speckles caused by global illumination. Thus, correspondence matching can be achieved robustly by the proposed oblique sinusoidal pattern illumination mode, from which projection functions along one or multiple orientations are obtained. Correspondence point can be calculated by intersecting the lines that satisfy the local maximum constraint. The local slice extension (LSE) method, which involves a "coarse to fine" localization procedure, is introduced for highly efficient projective function capture. The LSE method utilizes the local visible property of projection functions, which indicates that projection functions have nonzero values in local region, and concentrates on observing this region to accelerate capture efficiency. Experimental results show that partial frequency capture can obtain satisfactory results.

Two kinds of global illuminations, namely inter-reflections and subsurface scattering (Fig. 1), are discussed. Camera pixels can receive light from multiple projector pixels in real applications. Thus, multiple peaks are usually observed in projection function(s), as shown in Fig. 2(a). In the present paper, Several optimization methods, which enable 3D reconstruction under global illumination with dozens of projected patterns, are introduced in the present paper.

This paper offers the following contributions.

1. pPSI, a photography method that provides a robust and efficient way to analyze the light transport behavior and enables separation of the influences of lights due to global illumination in 3D scanning, is presented, thereby achieving 3D reconstruction under global illumination.

2. The relationship between projection functions and LTC is theoretically demonstrated, and the oblique sinusoidal pattern illumination mode is proposed.

3. The proposed theoretical foundation of pPSI, where the local maximum constraint, which provides basics that LTC and projection functions have equivalence class considering correspondence, is proven mathematically.

4. We introduce LES method, which is a highly efficient projection functions capture method, and the perfect reconstruction property for LSE is proven mathematically.

5. Optimization is conducted for pPSI. The number of projection functions required for pPSI for several situations is optimized, and the influence of capture ratio in LSE on the accuracy of the correspondence matching points is investigated. 3D reconstruction under global illumination can be achieved with dozens of projected patterns (<1 s capture time for the current system).

## 2 RELATED WORK

### 2.1 3D Reconstruction under Global Illumination

Several methods are developed to solve 3D reconstruction under global illuminations; these methods include high-frequency projection methods [5],[6],[7], regional projection methods [8],[9],[10],[11], and polarization projection methods [12],[13]. However, these methods are based on specific assumptions that may remain unsatisfied in real applications. For example, high-frequency projection methods, such as modulated phase-shifting [6] and micro-phase shifting [7], are mainly used to suppress low-frequency inter-reflections. These methods utilized high-frequency patterns and assume the existence of only low-frequency global illumination, which fails in practical situations. For strong subsurface scattering, the modulations of the projected high-frequency patterns are too low for high accuracy data reconstruction. Regional projection methods require previous information for the measured objects[8] or several rounds of projection mode [9],[10],[11] to suppress the occurrence of inter-reflections between surfaces. In addition, this method cannot suppress high-order inter-reflections. Polarization projection methods assume that direct illumination is not depolarized, which is not the case for diffuse objects. O'Tool *et al.* [15] introduced structured light transport (SLT) method for 3D reconstruction under global illumination. However, non-epipolar assumption was made in SLT, which assumes weak epipolar indirect illumination. This assumption can be broken in numerous real-world applications under strong epipolar plane reflection. On the contrary, pPSI makes no explicit assumption and handles inter-reflections (especially specular inter-reflections) and strong subsurface scattering simultaneously.

Several 3D reconstruction methods that assume no explicit assumption to overcome global illumination are recently introduced. Park *et al.* [14] proposed multipeak range imaging, wherein a single projector stripe line is illuminated at a time. However, this method requires long capture time because each projector stripe line has to be illuminated in turn. Each camera pixel in pPSI is treated as an independent unit and reconstructs a 1D projection function. Each measured pixel value has whole information of projection function. Thus, the excellent compressive property of Fourier single-pixel imaging can be explored, and the projection number can be markedly reduced. Diezu *et al.* [16] recently proposed a method called frequency shift triangulation. This method requires a calibration process to determine the minimal phase step of the measurement system and uses a dynamic programming method to eliminate erroneous data for successful 3D reconstruction. Zhang *et al.* [17] introduced a general mathematical model to solve 3D reconstruction under global illumination. Zhang *et al.* [18] later introduced a sparse multipath correction method that uses an application-specific Bayesian learning approach. This method requires an iterative optimization process, which can prolong the calculation time and is unsuitable for parallel computing. On the contrary, pPSI requires no additional calibration stage and the reconstruction algorithm requires no iterative process, which is suitable for parallel computing. Mixed peak phenomenon in Section 5.1, which cannot be solved by existing methods, is also introduced in this paper.

## 2.2 Light Transport Coefficients Capture

Light transport is important for computer vision and graphics. Debevec *et al.* [27] introduced the capture of a simplified 4D light transport function by a light stage. Masselus *et al.* [28] proposed the use of a projector–camera system to capture a 6D slice of the full light transport function. These early methods directly capture LTC, which results in a relatively low capture speed.

Adaptive methods, such as dual photography [29] and symmetric photography [30], as well as compressive imaging methods [30]-[33], are introduced for highly efficient light transport capture. However, these methods either require a complex illumination mode or a complex reconstruction algorithm. Primal–dual coding [34] and structured transport [15] are developed, wherein the illumination and camera pixels are controlled simultaneously to manipulate different components in the light transport between the projector and the camera. However, this method requires special optical design and hardware.

In our earlier researches, we proposed a single-pixel imaging method for LTC capture [19],[20]. Parallel single-pixel imaging (PSI) [21] was later introduced for efficient LTC capture using the local region extension (LRE) method. A compressive PSI [22] and deep-learning based PSI [23] are also introduced for highly efficient LTC capture. PSI is currently extended to paraxial systems [24] and for separating high-order inter-reflections [25].

The present paper aims to achieve a robust and efficient correspondence matching under strong global illumination for 3D scanning by introducing pPSI to reduce the capture of 4D LTC to projection functions. The most outstanding feature of pPSI lies in its capability to provide good balance between robustness and efficiency in 3D reconstruction under global illumination. The robustness indicates that pPSI makes no explicit assumption and handles inter-reflections (specular) and strong subsurface scattering simultaneously. The efficiency means that pPSI captures projection functions rather than LTC. Thus, pPSI is more efficient than the methods that capture LTC to solve 3D reconstruction under global illumination.

## 3 PARALLEL SINGLE-PIXEL IMAGING FOR LIGHT TRANSPORT CAPTURE

PSI captures LTC $h(u',v';u,v)$, which are a 4D dataset, between projector pixel $(u',v')$ and camera pixel $(u,v)$. LTC describe the image formation process, which is expressed as

$$I(u,v) = O(u,v) + \sum_{v'=0}^{N-1}\sum_{u'=0}^{M-1} h(u',v';u,v)P(u',v'), \qquad (1)$$

where $I(u,v)$ is the radiance captured by camera pixel $(u,v)$, $O(u,v)$ is the environment illumination, and $P(u',v')$ is the illuminated radiance of projector pixel $(u',v')$. $M$ and $N$ are the horizontal and vertical resolutions of the projector, respectively.

Jiang *et al.* [21] introduced the LRE method to accelerate the capture efficiency of PSI. This method assumes that the visible region of each pixel is confined in a local region, proving the perfect reconstruction property of LRE. Reference [21] provides detailed information on PSI. The visible region is referred in the present paper as *reception field*.

## 4 PROJECTIVE PARALLEL SINGLE-PIXEL IMAGING FOR EFFICIENT SEPARATION OF DIRECT AND GLOBAL ILLUMINATION

### 4.1 Local Maximum Constraint Proposition

This section provides the basics for obtaining direct illumination point (correspondence matching point) via projection functions. PSI requires complete LTC capture. However, in the case of 3D reconstruction, LTC contain over-complete information because only the direct correspondence point is extracted and stored for each PTI. The key observation underlying pPSI is that the direct illumination point can be recovered if the 1D projection function(s) of the 2D pixel transport image is captured [Fig. 2(a)]

$$f_\theta(\rho; u, v) = \Re_\theta[h(u', v'; u, v)]$$
$$= \sum_{v'=0}^{N-1} \sum_{u'=0}^{M-1} \cdot h(u', v'; u, v) \cdot \delta(\rho - u' \cos\theta - v' \sin\theta), \quad (2)$$

where $\Re_\theta[h(u', v'; u, v)]$ is the discrete Radon transform of LTC along direction $\theta$, which is the angle between the integral direction of the projection function and horizontal axis, $\rho$ is the coordinate of the projection function, and $\delta(\bullet)$ is the Dirac delt function. When each camera pixel is considered, $f_\theta(\rho; u, v)$ forms a 3D data cube.

Angle $\theta$ defines the *direction line* to which the pixel transport image is projected. The direction line is ob-tained through counter-clockwise rotation of the horizontal axis by $\theta$, as shown in Fig. 2(a). For example, vertical patterns correspond to the projection of pixel transport image to the horizontal axis. Thus, the angle $\theta=0$ in this case. Given a direction line and a point $(u', v')$, *projection line* is defined as the line passing through point $(u', v')$ and vertical to the direction line. Thus, the projection position of $(u', v')$ to the direction line is the intersection of the projection and direction lines, as shown in Fig. 2(a).

The fundamental principle for recovering direct illumination points given the projection functions lies in the local maximum constraint proposition, which provides constraint for the location of the correspondence matching point (direct illumination point) in the pixel transport image [Fig. 2(b)].

**Local Maximum Constraint Proposition.** *If the corresponding projection line does not pass through any speckles due to global illumination, then the direct illumination point on the pixel transport image is a local maximum point on the projection functions.*

**Proof** of Local Maximum Constraint Proposition can be found in Appendix A. □

Local maximum constraint proposition provides a necessary condition for the location of correspondence matching points. Figures 2 (a) and (b) provide an intuitive explanation of local maximum constraint proposition.

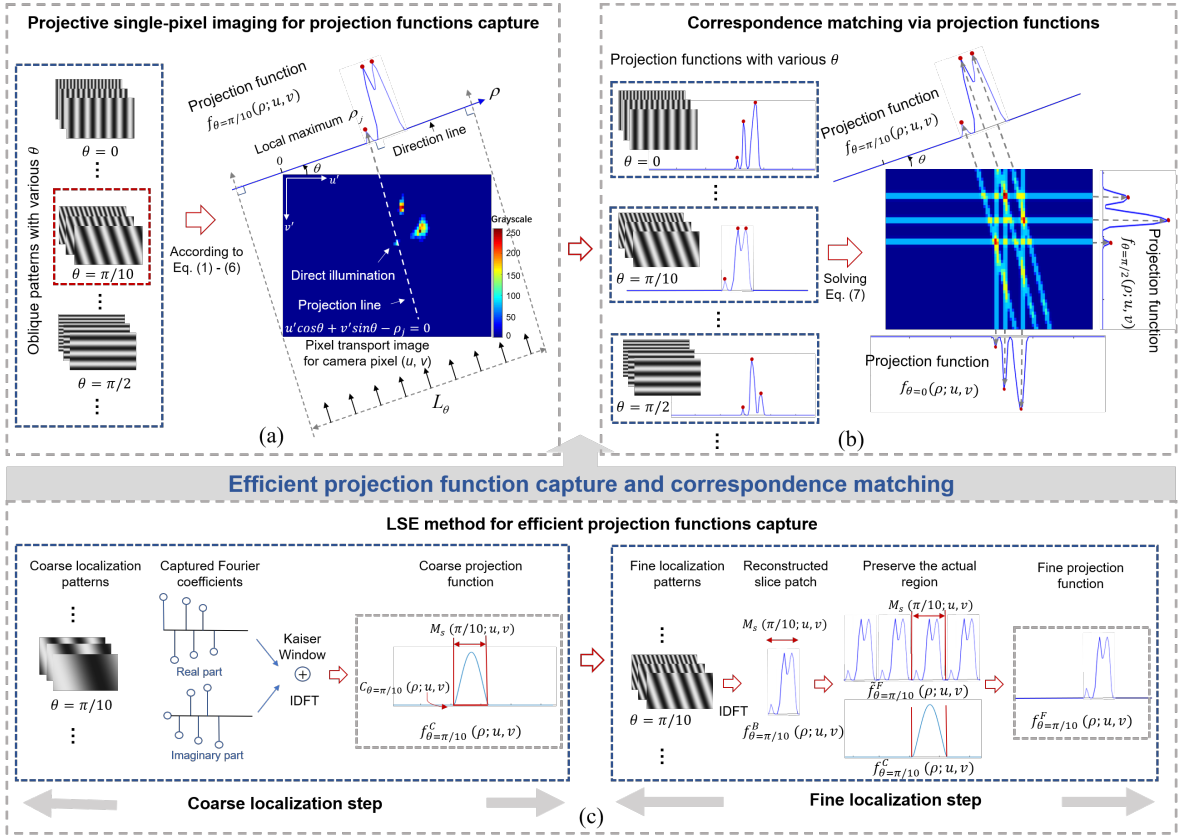

Fig. 2. Fundamental principles of pPSI. (a) Projective single-pixel imaging for projection functions capture. The illumination of oblique patterns is equivalent to the application of Radon transform to PTI. The correspondence matching points of pixel transport image are retained in the projection functions (red spots). (b) Correspondence matching via projection functions. The red spots are local maximum. (c) LSE method for efficient projection functions capture.

Suppose multiple projection functions $f_\theta(\rho; u, v)$ with $D$ directions are obtained, with the direction of $d$-th projection at $\theta_d$. The $d$-th projection function has a total of $T_d$ local maxima. The $j$-th local maximum of the $d$-th projection function is denoted as $\rho_d^j$. A grayscale centroid subpixel matching processing is introduced in [19] and [20], which should be applied to obtain $\rho_d^j$. A linear equation can be formed if any local maximum is chosen from each projection line, and each such combination is intersected according to Eq. (3):

$$\begin{pmatrix} \cos\theta_1 & \sin\theta_1 & -\rho_1^m \\ \cos\theta_2 & \sin\theta_2 & -\rho_2^n \\ \vdots \\ \cos\theta_D & \sin\theta_D & -\rho_D^p \end{pmatrix} \begin{pmatrix} u' \\ v' \\ 1 \end{pmatrix} = \begin{pmatrix} 0 \\ 0 \\ \vdots \\ 0 \end{pmatrix}, \qquad (3)$$

where $m, n, p$ are integers, taking any combination that satisfy $m \in [0, T_1)$, $n \in [0, T_2)$, and $p \in [0, T_D)$, respectively. These linear equations can be solved by singular value decomposition (SVD), and unintersected combinations are eliminated by checking the rank. The intersected points caused by global illuminations are preserved and are then eliminated by epipolar constraint between the projector and camera, as conducted in [19]-[25].

### 4.2 Projective Single-pixel Imaging for Projection Functions Capture

This section shows that illuminating oblique sinusoidal patterns is equivalent to applying Radon transform to the pixel transport image. Fig. 2 (a) provides the basic idea.

The oblique S-step ($S \geq 3$) sinusoidal patterns with the following form are illuminated

$$P_i(u', v'; k, \theta) = a + b \cdot \cos[\frac{2\pi k}{L_\theta}(u'\cos\theta + v'\sin\theta) + \frac{2\pi i}{S}], \qquad (4)$$

where $i$ denotes the phase step, and take values of $i = 0, 1, \dots S-1$. $a$ and $b$ are the average and contrast of the patterns, respectively. $k$ is the discrete frequency samples and take values of $k = 0, 1 \dots K$ and $K \leq L_\theta - 1$. $L_\theta$ is the equivalent projector resolution in the projector range for the directional projection function with an angle of $\theta$ [Fig. 2(a)], which can be calculated as

$$L_\theta = \begin{cases} \lceil M \cdot \cos\theta + N \cdot \sin\theta \rceil & 0 \leq \theta \leq \pi/2 \\ \lceil -M \cdot \cos\theta + N \cdot \sin\theta \rceil & \pi/2 < \theta < \pi \end{cases}, \qquad (5)$$

where $\lceil \cdot \rceil$ is the ceiling function.

According to Eq. (1), the captured intensity for camera pixel $(u, v)$ can be calculated as

$$I_i(u, v; k, \theta) = O(u, v) + \sum_{v'=0}^{N-1}\sum_{u'=0}^{M-1} a \cdot h(u', v'; u, v) \\ + \sum_{v'=0}^{N-1}\sum_{u'=0}^{M-1} b \cdot h(u', v'; u, v) \cdot \cos[\frac{2\pi k}{L_\theta}(u'\cos\theta + v'\sin\theta) + \frac{2\pi i}{S}]. \qquad (6)$$

Suppose that all of phase step $i$ is captured. Given a frequency sample $k$ and direction $\theta$, the following quantity can be obtained:

$$F_\theta(k; u, v)$$
$$= \sum_{i=0}^{S-1} I_i(u, v; k, \theta)\cos(2\pi i/S) + j\sum_{i=0}^{S-1} I_i(u, v; k, \theta)\sin(2\pi i/S) \qquad (7)$$
$$= \frac{S}{2} \cdot \sum_{v'=0}^{N-1}\sum_{u'=0}^{M-1} b \cdot h(u', v'; u, v) \cdot \exp[-\frac{2\pi k}{L_\theta}(u'\cos\theta + v'\sin\theta)].$$

Eq. (7) holds due to Lagrange's trigonometric identities. The product-to-sum formulas of trigonometric identities and the Euler's formula should also be applied.

When patterns with $k = 0, 1 \dots L_\theta - 1$ are illuminated and $F_\theta(k; u, v)$ are calculated as Eq. (7), the projection function $f_\theta(\rho; u, v)$ shown in Fig. 2 can then be obtained by taking IDFT to $F_\theta(k; u, v)$.

$$f_\theta(\rho; u, v)$$
$$= IDFT\{\frac{Sb}{2} \cdot \sum_{v'=0}^{N-1}\sum_{u'=0}^{M-1} \cdot h(u', v'; u, v) \cdot \exp[-\frac{2\pi k}{L_\theta}(u'\cos\theta + v'\sin\theta)]\}$$
$$= \frac{Sb}{2} \cdot \sum_{r=-\infty}^{+\infty}\sum_{v'=0}^{N-1}\sum_{u'=0}^{M-1} \cdot h(u', v'; u, v) \cdot \delta(\rho - u'\cos\theta - v'\sin\theta - rL_\theta)$$
$$= \frac{Sb}{2} \cdot \Re_\theta[h(u', v'; u, v)], \qquad (8)$$

where $r$ are integers. $f_\theta(\rho; u, v)$ contains an infinite sum term; however, the pixel transport image has nonzero values only in one continuous region with a length of $L_\theta$. Thus, Eq. (8) precisely applies discrete Radon transform to the pixel transport image along direction $\theta$, with a scale factor. Three-step sinusoidal oblique patterns are used in the present paper for the capture of projection functions because these patterns are standard in the field of SLS.

### 4.3 LSE Method for Efficient Projection Function Capture

The LSE method, which is implemented by a "coarse to fine" localization procedure, is introduced for highly efficient projection function capture [Fig. 2(c)]. The fundamental basis of LSE can be proven by reducing the LRE reconstruction theorem [21] to a 1D case (refer to Appendix B for detailed information). Compared with the LRE method, the LSE method captures projection functions with different orientations. Thus, in the LSE method, concepts equivalent to the size and location of the reception field in the LRE method are the size and location of the reception field projected along projection direction with $\theta$. These concepts are referred to as the size and location of $\theta$ projected reception field [Fig. 2(c)]. The accelerated projective single-pixel imaging of the LSE is referred to as pPSI.

**Coarse localization step.** This step has a two-fold goal: detecting and obtaining the coarse location and size of the projected reception field. Oblique patterns with the form of Eq. (4) are projected. The obtained intensities are arranged as Eq. (7), and 1D IDFT is applied to the resulting quantities. A Kaiser window [35], wherein the shape parameter $\beta$ is set as 5, is applied on the tested low-frequency samples to eliminate ringing effect. Coarse projection functions $f_\theta^C(\rho; u, v)$ can then be obtained. The coarse location of the projected reception field $C_\theta(\rho; u, v)$ can be determined; this location is a mask that has a value of one when the reconstructed coarse projection functions are larger than the noise threshold and zero otherwise. The size of the projected reception field $M_s(\theta; u, v)$ is determined between the length in the first and last positions which is larger than the noise threshold. The number of frequencies for coarse localization is set as 10 in the pre-sent paper. Appendix C is used as reference for a theoretical analysis of the relationship between localization accuracy and frequency number in the coarse localization step and how the number of frequencies for coarse localization is chosen.

**Fine localization step.** This step comprises three substeps: the projection of fine location patterns, the reconstruction of fine projection function, and the merging of coarse and fine steps. The fine projection patterns are the 1D case of the periodic extension patterns introduced in reference [21].

First, the fine location patterns with the following form are projected as follows:

$$\tilde{P}(u',v';k,\theta) = a + b \cdot \cos[\frac{2\pi k_\theta}{M_\theta}(u'\cos\theta + v'\sin\theta) + \frac{2\pi i}{S}], \quad (9)$$

where $\theta$ is the angle between the integral direction of the projection function and horizontal axis. $i$ denotes the phase step and take values of $i = 0, 1, \ldots S - 1$. $a$ and $b$ are the average and contrast of the patterns, respectively. $k$ is the discrete frequency samples and takes the value of $k_\theta = 0, 1 \ldots M_\theta - 1$. $M_\theta$ is the size of the maximum of $\theta$ projected reception field for each camera pixel and is defined as follows:

$$M_\theta = \max_{(u,v)}[M_s(\theta; u, v)]. \quad (10)$$

Second, the captured intensities when each pattern is projected are arranged as Eq. (7), and 1D IDFT is applied to reconstruct slice patch $f_\theta^B(\rho; u, v)$. This reconstructed slice patch is then extended periodically for the projection functions with resolution of $L_\theta$ and can be expressed as follows:

$$\tilde{f}_\theta^F(\rho; u, v) = \sum_{r=0}^{\left\lceil \frac{L_\theta}{M_\theta} \right\rceil} f_\theta^B(\rho - rM_\theta; u, v), \quad (11)$$

where $r$ are integers, and $\rho = 0, 1 \ldots L_\theta - 1$.

Finally, the fine projection functions are reconstructed by merging the information from coarse and fine steps,

that is, by preserving the nonzero region of $C_\theta(\rho; u, v)$ obtained from coarse localization step, as expressed by

$$f_\theta^F(\rho; u, v) = \tilde{f}_\theta^F(\rho; u, v) \cdot C_\theta(\rho; u, v), \quad (12)$$

where $\cdot$ denotes the element-wise product.

**Analysis on the improvement in capture efficiency.** Compared with PSI, the capture complexity of pPSI is reduced from $O(M_s^2)$ to $O(M_s)$, where $M_s$ is the size of the reception field.

# 5 OPTIMIZATION ON PROJECTIVE PARALLEL SINGLE-PIXEL IMAGING

Optimization strategies for pPSI are proposed on the basis of the theoretical foundation in Section 4. The optimization strategies comprise two aspects: optimization on the number of projections (Sections 5.1 and 5.2) and on the partial frequency captured (Section 5.3).

## 5.1 Projective Parallel Single-pixel Imaging by Four Projections Strategy

A RANSAC-based intersection method with four projections is introduced in this subsection to calculate the corresponding points robustly for inter-reflection scenes. First, mixed peaks will be introduced to explain the unsuitability of direct intersection with four projections in scenes dominated by inter-reflections.

**Mixed peaks in inter-reflections.** Mixed peaks refer to the phenomenon in which multiple peaks are mixed into one peak due to improper selection of the directions of projection functions. An illustration is provided in Fig. 3(a). Three speckles are observed on the LTC due to inter-reflections. Projection functions 1–3 demonstrate three peaks. However, projection function 4 only has two peaks due to its improper direction. Under careful examination, peak $\alpha$ is caused by speckle $\beta$ and $\gamma$ simultaneously, which corresponds to the situation when local maximum constraint proposition is violated. Thus, if the corre-sponding points are calculated via direct intersection with four projections, then large error or missing data can be incurred, refer to Figs. 3(b) and 3 (g).

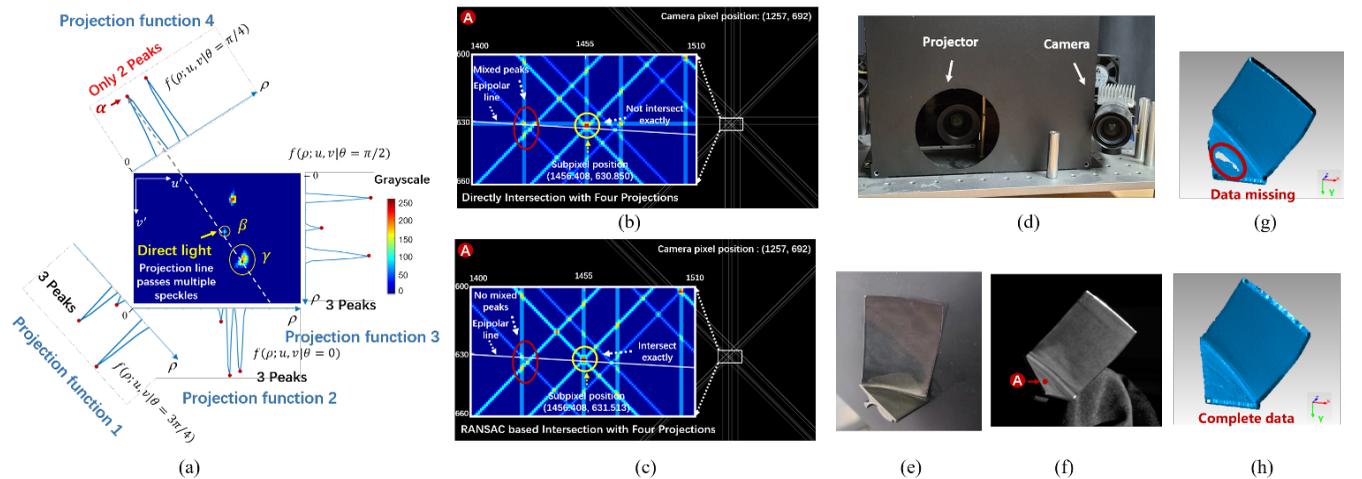

Fig. 3. Strategy of four projections for projective parallel single-pixel Imaging. (a) Mixed peaks for inter-reflection scenes. (b) Direct intersection with four projections when mixed peaks were incurred. (c) RANSAC-based intersection method with four projections when mixed peaks were incurred. (d) 3D imaging system used in this paper. (e) Metal blade. (f) Metal blade from the camera perspective, where the pixel denoted with A corresponds to the position where the process of back-projection and intersection is shown in subfigures (b) and (c). (g) The 3D data reconstructed via direct intersection with four projections. Data are incomplete because of mixed peaks. (h) The 3D data reconstructed by RANSAC-based intersection method with four projections. Complete data are obtained.

**Random Sample Consensus (RANSAC)-based intersection method with four projections.** A RANSAC-based intersection method is introduced to overcome the failure cases in the presence of mixed peaks. This method first checks the consensus of the intersection and then eliminates the inconsistent peaks. Appendix D provides detailed information of this method.

The 3D imaging system shown in Fig. 3(d) is used to test the effectiveness of this method and reconstruct 3D data of a metal blade shown in Fig. 3(e). Section 6 presents detailed imaging parameters of the system. Oblique patterns with four orientations of $\theta = 0°, 45°, 90°, 135°$ were projected. The length of the reception field for each direction was 100 pixels according to the coarse localization step.

Figs. 3 (b) and (c) correspond to the process of back-projection and intersection of pixel A shown in Fig. 3(f), by direct intersection and RANSAC-based method, respectively. Mixed peaks in Fig. 3(b) occurred because the horizontal back-projection line simultaneously passed through the direct speckle (denoted by yellow circle) and another speckle caused by inter-reflections (denoted by red circle). The calculated corresponding position was (1456.408, 630.850). The 3D points via direct intersection with four projections are shown in Fig. 3 (g). Missing data were incurred because of the large error in calculating the corresponding points (even failure to pass the epipolar constraint). The horizontal back-projection line in Fig. 3 (c) had been eliminated by the RANSAC- based intersection method. The calculated corresponding position was (1456.408, 631.513), and the difference was more than 0.6 pixels in vertical direction. The 3D points by the RANSAC-based intersection method are shown in Fig. 3 (h), which implies the effectiveness of the proposed method.

### 5.2 Quick Projective Parallel Single-Pixel Imaging by Unidirectional Projection Strategy

A quick pPSI, which calculates corresponding points via unidirectional projection strategy, is proposed to further improve the capture efficiency. The number of projection functions is optimized to its minimum (Fig. 4) via the aforementioned strategy. The calculation of 3D points contains two sub-steps: calculation of candidate matching points and elimination of the virtually matched 3D points. This strategy can be used only under such situations when the mixed peaks is guaranteed not to occur due to viewpoint optimization.

**Calculation of candidate matching points.** Projection function with only one direction is captured. Thus, the candidate matching points are calculated by the intersection of the back-projected local maximum on the projection function with the epipolar line, as shown in Fig. 4 (a). Virtual matching points that cause virtually matched 3D points are also observed, as shown in Fig.4(a).

Suppose that the projection function with angle $\theta$ is obtained, $\rho_j$ is the $j$-th local maximum on the projection function, and $a, b, c$ are the parameters for the epipolar line in its standard form. Candidate matching points are calculated by solving the following linear equation

$$\begin{pmatrix} \cos\theta & \sin\theta & -\rho_j \\ a & b & c \end{pmatrix} \begin{pmatrix} u' \\ v' \\ 1 \end{pmatrix} = \begin{pmatrix} 0 \\ 0 \\ 0 \end{pmatrix}. \tag{13}$$

The camera and the projector are arranged horizontally for the established 3D imaging system in this paper [Fig. 3(d)]; thus, the epipolar line is close to a horizontal line [Fig. 4 (a)]. Obtaining projection function with direction $\theta = 0°$ leads to satisfactory results.

**Elimination of the virtually matched 3D points.** An algorithm that adopts continuity constraint is used to eliminate the virtually matched 3D points [Fig. 4(b)]. Appendix E provides detailed information regarding this algorithm.

### 5.3 Partial Frequency Capture in Fine Localization Step

Partial frequencies in Eq. (9), where the projected pattern frequencies can be confined within a partial low-frequency range, are captured to maximize the capture efficiency. For example, a certain percentage of low-frequency patterns generated by Eq. (9) can be projected. Experimental results for inter-reflections and subsurface scattering are provided to study the influence of the percentage to the subpixel accuracy of the local maximum in projection functions. The capability of 3D reconstruction using only partial frequencies in pPSI is attributed to the excellent compression performance of Fourier spectrum. The subpixel accuracy of the local maximum in projection functions is measured by subpixel matching error (SME),

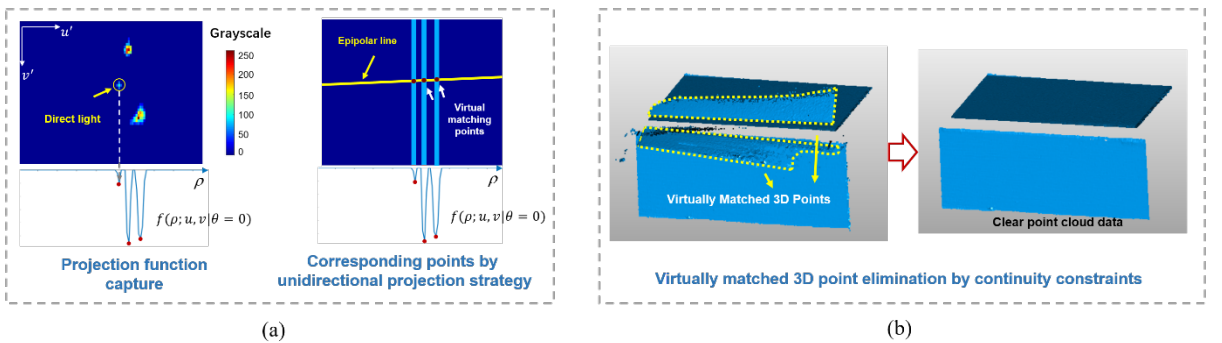

(a)

(b)

Fig. 4. 3D reconstruction by unidirectional projection strategy. (a) Calculation of candidate matching points by unidirectional projection strategy. (b) Virtually matched 3D point elimination by continuity constraints.

which is defined by

$$SME(\boldsymbol{u}_R, \boldsymbol{u}_P) = \frac{1}{2}[(u'_R - u'_P)^2 + (v'_R - v'_P)^2], \qquad (14)$$

where $\boldsymbol{u}'_R = (u'_R, v'_R)$ and $\boldsymbol{u}'_P = (u'_P, v'_P)$ are the subpixel matched points obtained by PSI and pPSI, respectively, demonstrating varying capture ratios. The tested capture ratios were as follows: 5%, 8%, 10%, 12%, 14%,16%, 18%, 20%, 22% 25%,30%, 35%, 40%, 50%, 60%, 70%, 80%, and 90%,100%.

Normalized energy distribution (NED) is used to explain the efficiency of partial frequency capture. NED is defined as follows:

$$NED[F^s(k_s)] = \frac{1}{P_N} \sum_{(u,v)} \left| F^s(k_s; u, v) \right| \qquad (15)$$

where $P_N$ is the normalization factor, which is calculated by

$$P_N = \max_{k_s}[\sum_{(u,v)} \left| F^s(k_s; u, v) \right|] \qquad (16)$$

where $k_s = 1, 2, \ldots \lfloor U_s / 2 \rfloor - 1$.

**Partial frequency capture for inter-reflections.** A triangular groove [Fig. 5 (a)] comprising metal is used to analyze the influence of partial capture on SME for scenes dominated by inter-reflections. PSI and pPSI implemented by unidirectional projection strategy (Section 5.2) were performed, and the SMEs were calculated for pixels inside the shaded region shown in Fig. 5 (b). The averaged SME of the triangular groove is plotted in Fig. 5 (c). The figure reveals that 40% capture ratio point (represented by a red star) provides the best tradeoff between accuracy and efficiency because the gradient of the curve markedly changes near this point. The curve rapidly drops before the point and then slowly descents after such point. The average SME of 40% capture is 0.255 pixel, which is almost the same as the SME value 0.262 of 100% capture. Thus, using the capture ratio of 40% provides satisfactory results.

The NED is presented in Fig. 5(d) to address the possibility of using 40% capture ratio to obtain satisfactory results. The energy of the first 40% frequencies account for 90.36% of the total energy. Thus, only a substantially small amount of information will be lost under partial low-frequency capture introduced in this section.

**Partial frequency capture for subsurface scattering.** A candle [Fig. 5 (e)] is used for scenes dominated by subsurface scattering to analyze the influence of partial capture on SME. The pixels inside the shaded region shown in Fig. 5(f) were calculated. Similar to the situation dominated by inter-reflections, the averaged SME is plotted in Fig. 5 (g). The figure shows that the 16% capture ratio point (represented by a red star), which is lower than the inter-reflection counterpart, provides the best tradeoff between accuracy and efficiency. The averaged SME of 16% capture is 0.482 pixel. The NED is also shown in Fig. 5(h), where the energy of the first 16% frequencies accounts for 95.06% of the total energy. A low-frequency capture ratio can be applied for subsurface scattering because of the low-pass characteristics of translucent objects.

## 6 EXPERIMENTS AND EVALUATIONS

The experimental setup comprises a camera and a projector, as shown in Fig. 3(d). The resolutions of the camera and projector are 1600 × 1200 and 1920 × 1080, respectively. The frame rate of the projector is synchronized with that of the camera. The capture rate of the system was 165 frames per second (fps). Several challenging scenarios were validated. Sections 6.1–6.3 used the RANSAC-based intersection method in Section 5.1, and Sections 6.4–6.5 used the unidirectional projection strategy in Section 5.2.

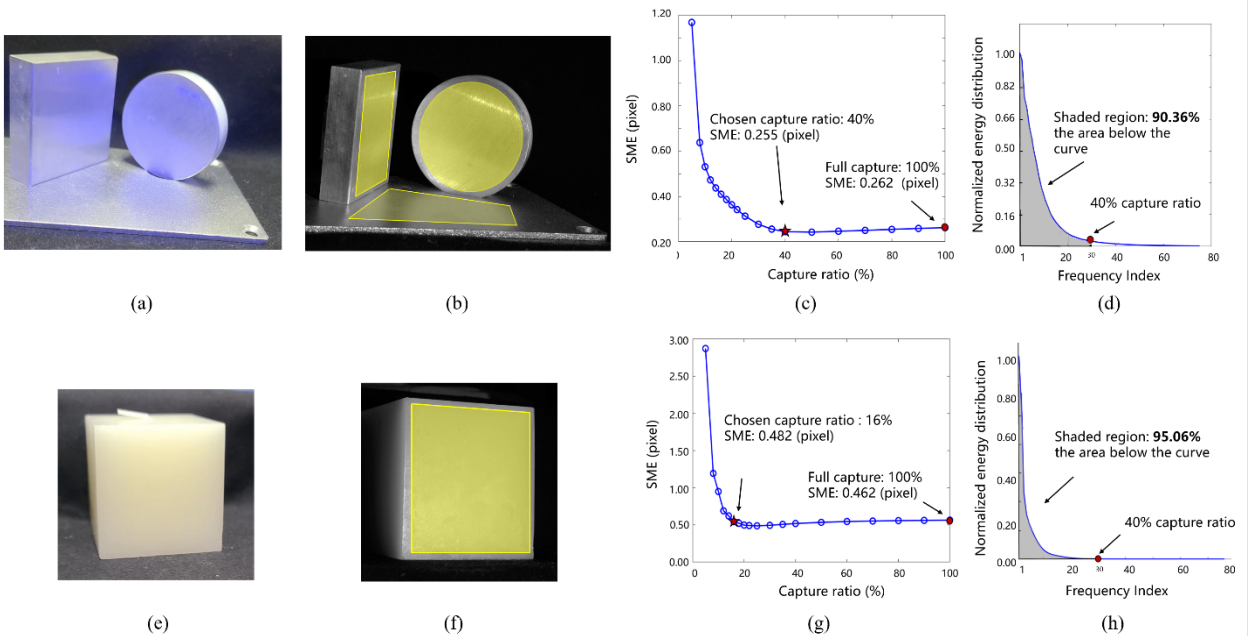

Fig. 5. Analysis of the influence of the capture ratio to the subpixel accuracy of the local maximum in projection functions. (a) Triangular groove made by metal. (b) The calculation region for the triangular groove. (c) The influence of capture ratio on SME for the triangular groove. (d) Normalized energy distribution with respect to frequency for the triangular groove. (e) Candle. (f) The calculation region for the candle. (g) The influence of capture ratio on SME for the candle. (h) Normalized energy distribution with respect to frequency for the candle.

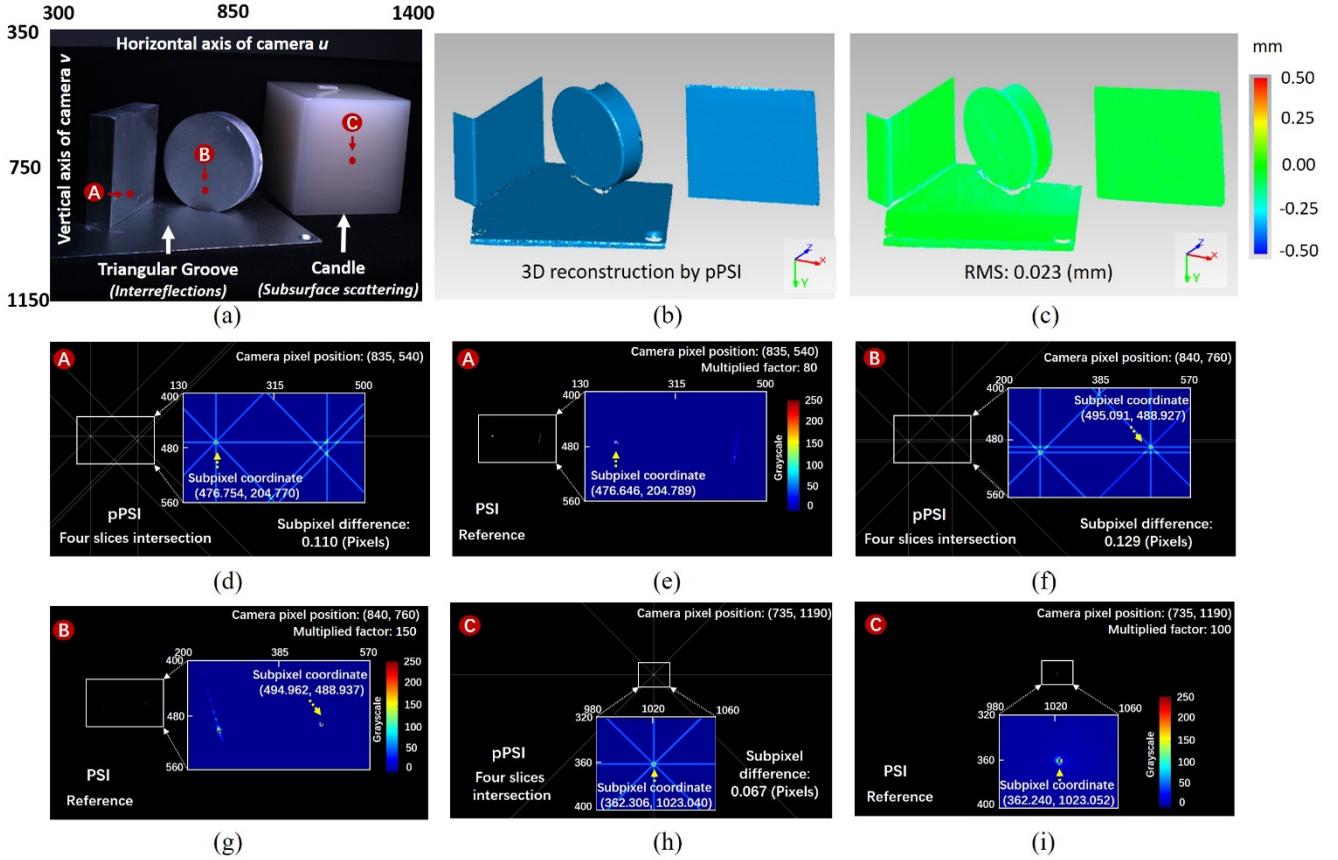

Fig. 6. Comparison between pPSI and PSI using the compound scene. (a) The compound scene contains inter-reflections and subsurface scattering. The camera coordinates are depicted. The positions of three points, namely points A, B, and C, are indicated by red circles. The intersection positions by pPSI and LTC by PSI of the three points are shown in (d)–(i). (b) 3D shape reconstructed by pPSI. (c) Error map of the point cloud data between pPSI and PSI. RMS was 0.023 mm. (d), (f), and (h) are the intersection points calculated by pPSI. The camera positions are indicated on the upper right corner. On the upper left corner of each of these subfigures, a circle with a letter inside indicates the corresponding point to the subfigure. (e), (g), and (i) are the light transport coefficients and the subpixel matched positions calculated by PSI. The difference of the subpixel matched positions between pPSI and PSI are shown in (d), (f). and (h).

## 6.1 Compound Scene

A compound scene, which contains a triangular groove and a candle, is used to compare pPSI and PSI. The image of the investigated scene is shown in Fig. 6(a). Oblique patterns with four orientations of $\theta = 0°, 45°, 90°, 135°$ are projected. The length of the reception field for each direction is 150 pixels according to the coarse localization step, which results in a total number of 336 patterns by pPSI (25% capture ratio for fine localization leads to satisfactory results). The calculation of pattern number required by pPSI are provided in the Appendix F. Total acquisition time was 2 s. The reconstructed 3D shape is shown in Fig. 6(b). The number of patterns required by PSI is 51,000. Total acquisition time is approximately 5 min. Thus, pPSI provides approximately 150-fold improvement in the present experiment. The error map between pPSI and PSI is shown in Fig. 6(c). The root-mean-square (RMS) error is 0.023 (mm). This experiment illustrates that pPSI is efficient and robust for 3D reconstruction in situations dominated by global illumination.

The LTC for three typical points are shown in Figs. 6(e), (g) and (i). The coordinate of each correspondence point is also shown. We provided the coordinate of each correspondence point calculated by pPSI in Figs. 6 (d), (f) and (h). These correspondence points are calculated as intersection points of the projection lines, as shown in Figs. 6 (d), (f) and (h). The differences of these correspondence point calculated by pPSI and PSI are also shown.

## 6.2 Inter-reflections

In this subsection, pPSI is tested in situations dominated by inter-reflections. In the first scene, a gypsum bear was placed near a mirror [Fig. 7(a)]. High-frequency inter-reflections result in overlapped patterns, which is challenging for FPP. In the second scene [Fig. 7(b)], two metal blades were measured. The specular reflection also incurs overlapped patterns, which results in large data missing areas.

The measurement parameters are the same as that in Section 6.1. The number of projected patterns is 336. The difference between pPSI and PSI is negligible, but pPSI achieved about 150-fold improvement in terms of acquisition time.

The accuracy of pPSI in situations dominated by inter-reflections is analyzed by a V-Groove that contains two metal gauge blocks [Fig. 8(a)]. A plane was separately fitted for the upper and lower planes. The RMS errors between the fitted planes and data points of the upper and lower planes were 0.021 (mm) and 0.015 (mm), respectively.

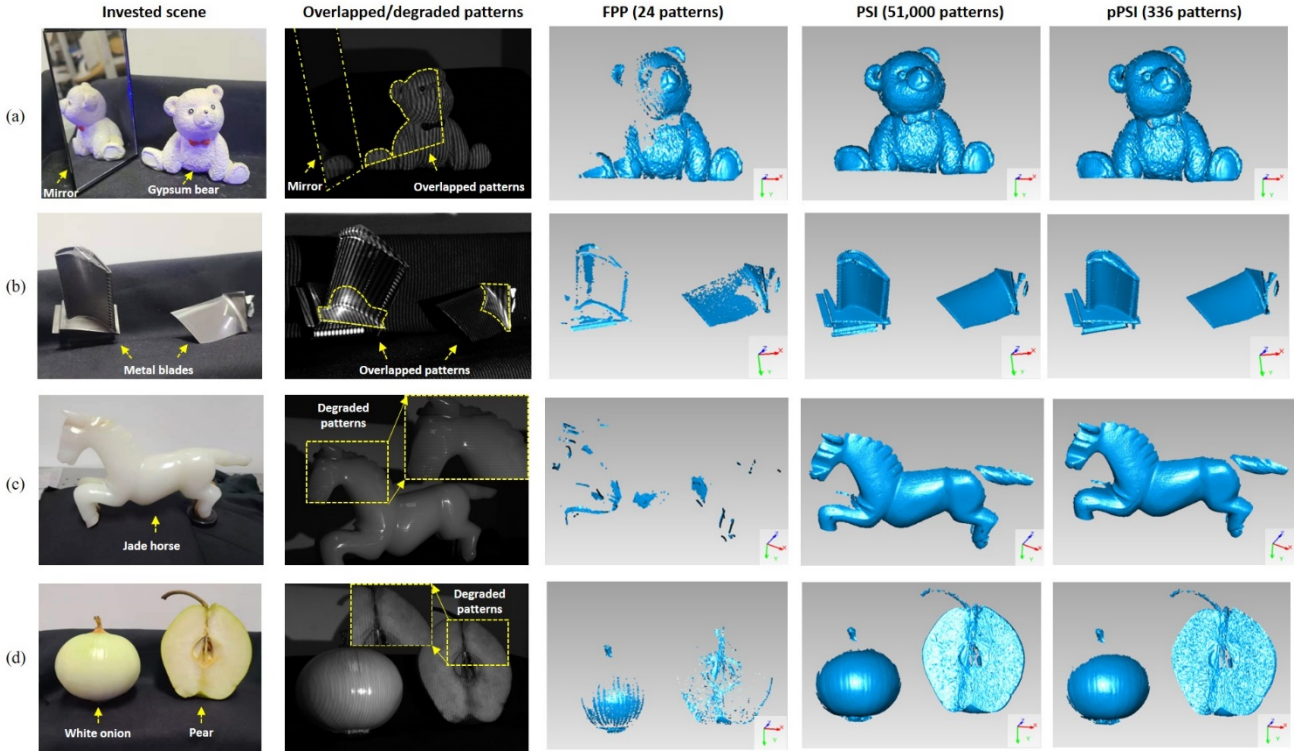

Fig. 7. Comparison of 3D shape reconstruction results among FPP, PSI and pPSI. The overlapped/degraded patterns are shown. (a) Gypsum bear. A mirror was placed near the bear such that high frequency strong inter-reflections dominate. (b) Metal blades. (c) Jade horse. (d) White onion and pear.

### 6.3 Subsurface Scattering

In this subsection, pPSI is tested in situations dominated by subsurface scattering. A jade horse was investigated in the first scene [Fig. 7(c)]. Strong subsurface scattering results in degraded patterns, which is challenging for FPP. A white onion and a pear were investigated in the second scene [Fig. 7(d)]. However, the FPP method still failed to reconstruct satisfactory 3D shape. pPSI and PSI can reconstruct high-quality 3D shapes for the two challenging scenes.

The measurement parameters are the same as that in Section 6.1. The number of projected patterns is 336. The difference between pPSI and PSI is negligible. However, pPSI achieved approximately 150-fold improvement considering acquisition time.

The accuracy of pPSI in situations dominated by subsurface scattering is analyzed by a polyamide sphere with diameter of 25.449 (mm) [Fig. 8(b)]. A sphere was fitted by the reconstructed points. The RMS error between the fitted sphere and the reconstructed data points was 0.031 (mm), and the diameter of the fitted sphere was 25.432 (mm). Thus, the absolute reconstruction error of pPSI was 0.017 mm, and the uncertainty of the measurement was 0.031 (mm).

### 6.4 Experimental Results of Unidirectional Projection Strategy

The effectiveness of unidirectional projection strategy is performed in this subsection. We refer to pPSI that adopts unidirectional projection strategy as quick pPSI.

**Inter-reflections.** The invested scenes are shown in Figs. 9 (a)–(c). The scenes include the gypsum bear with a mirror nearby, a metal part where high-frequency inter-reflections are dominated because of the glossy reflections caused by metal, and the V-groove comprising metal. High-frequency inter-reflections are dominant in these scenes.

Patterns with $\theta = 0°$ are projected. The length of the reception field for each direction was 150 pixels according to the coarse localization step, which results in a total number of 84 patterns (25% capture ratio is used in these experiments because it leads to satisfactory results; Sections 5.3 and 6.5 provide a detailed discussion on the influence of subpixel matching error and accuracy). Total acquisition time was 0.5 s. The reconstructed 3D shape is shown in Figs. 9 (a)–(c), where complete and high-quality 3D data

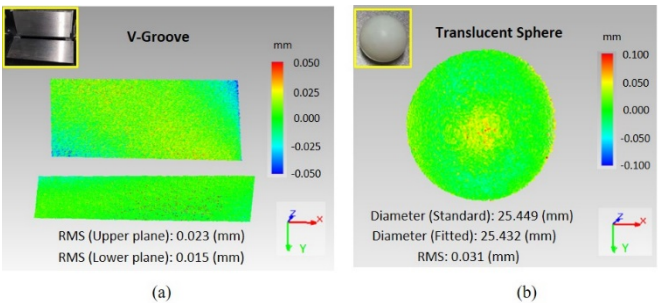

Fig. 8. Accuracy analysis of pPSI. (a) Accuracy analysis using V-Groove when inter-reflections are present. (b) Accuracy analysis using translucent sphere when subsurface scattering is present.

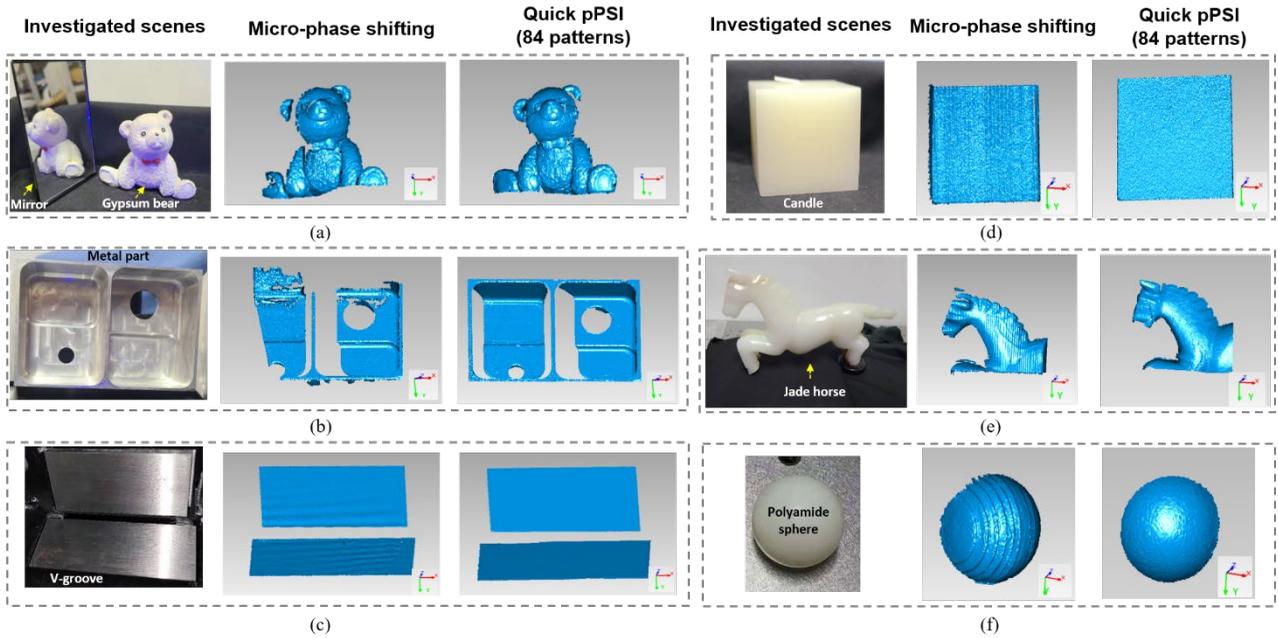

Fig. 9. Comparison with micro-phase shifting method. (a) Gypsum bear, where the mirror incurs high-frequency inter-reflections that micro-phase shifting method cannot handle. (b) Metal part, where the glossy reflection of metal incurs high-frequency inter-reflections that micro-phase shifting method cannot handle. (c) V-groove. Large errors were incurred for micro-phase shifting method. (d) Candle. (e) Jade horse. (f) Polyamide sphere. Micro-phase shifting method incurred large errors for scenes in subfigures (d)–(f), while pPSI reconstructed high-quality 3D data.

were obtained by quick pPSI (third column in each sub-figure).

**Subsurface scattering.** The invested scenes are shown in Figs. 9 (d)–(f). The scenes include a candle, a jade horse, and a polyamide sphere. Strong subsurface scattering was dominant in these scenes. The measurement parameters are the same as that of the inter-reflection scenes in Section 6.4. The number of projected patterns was 84. Total acquisition time was 0.5 s. The reconstructed 3D shape is shown in Figs. 9 (d)–(f), where complete and high-quality 3D data are obtained by quick pPSI (third column in each sub-figure).

### 6.5 Experimental Results of Partial Frequency Capture

The influence of capture ratio on measurement accuracy is investigated in this subsection under inter-reflections and subsurface scattering.

**Inter-reflections.** The V-groove [Fig. 9 (c)] was imaged by quick pPSI. The imaging parameters were the same as that in Section 6.4. The RMS value was calculated for various capture ratios. The experimental results are shown in Table 1. The experimental results reveal that the change of RMS is only 0.001 (mm) when the capture ratio changes from 25% to 100%. Overall, partial frequency capture introduced in Section 5.3 does not seriously affect the quality of point cloud. Only 84 patterns are required when using 25% capture ratio. The experiment also performed capture ratio experiments with less frequencies, such as 20% and 15%. However, accurate 3D data cannot be obtained when using these capture ratios. This finding is consistent with the result of Fig. 5 (c). Notably, superior RMS values are obtained by using a low capture ratio, which is reasonable.

The low-frequency capture method removes the noise in the original data and provides smooth results.

TABLE 1
INFLUENCE OF CAPTURE RATIO ON ACCURACY UNDER INTER-REFLECTIONS

| Capture ratio (%) | Number of projected patterns | Acquisition time under 165 fps capture rate | RMS of upper plane (mm) | RMS of lower plane (mm) |
|---|---|---|---|---|
| 25 | 84 | <1s | 0.025 | 0.016 |
| 30 | 96 | <1s | 0.025 | 0.016 |
| 35 | 108 | <1s | 0.025 | 0.016 |
| 40 | 117 | <1s | 0.026 | 0.017 |
| 50 | 141 | <1s | 0.026 | 0.017 |
| 80 | 210 | ≈1.5s | 0.026 | 0.017 |
| 100 | 255 | ≈1.7s | 0.026 | 0.017 |

TABLE 2
INFLUENCE OF CAPTURE RATIO ON ACCURACY UNDER SURFACE SCATTERING

| Capture ratio (%) | Number of projected patterns | Acquisition time under 165 fps capture rate | Diameter error (mm) | RMS (mm) |
|---|---|---|---|---|
| 10 | 35 | <0.5s | +0.063 | 0.069 |
| 15 | 60 | <0.5s | +0.053 | 0.057 |
| 20 | 72 | <1s | +0.034 | 0.048 |
| 25 | 84 | <1s | +0.025 | 0.043 |
| 30 | 96 | <1s | +0.021 | 0.039 |
| 35 | 108 | <1s | +0.022 | 0.036 |
| 40 | 117 | <1s | +0.022 | 0.035 |
| 50 | 141 | <1s | +0.020 | 0.034 |
| 80 | 210 | ≈1.5s | +0.021 | 0.036 |
| 100 | 255 | ≈1.7s | +0.020 | 0.036 |

**Subsurface scattering.** The polyamide sphere [Fig. 9 (f)] was imaged by quick pPSI. The imaging parameters were the same as that in Section 6.4. The calculated diameter error and RMS values for capture ratio vary from 10% to 100%, as shown in Table 2. The table reveals that the variation of diameter deviation is not obvious when the capture ratio changes from 30% to 100%. However, the diameter error and RMS increased rapidly with the decrease of the capture ratio when the capture ratio is 25% or less. Different from inter-reflections, partial frequency capture for subsurface scattering will not induce fatal errors under low capture ratio but will reduce the accuracy. Therefore, the capture ratio can be appropriately reduced for situations where high speed and low accuracy is required.

The results in Tables 1 and 2 reveal that 25% capture ratio is the lowest ratio that can handle inter-reflections and subsurface scattering simultaneously; therefore, 25% capture ratio was used for the experiment in Sections 6.1–6.4. Appendix F is used as reference for calculating the number of projected patterns.

## 6.6 Comparison to State-of-the-Art Methods

The proposed method is compared in this section with micro-phase shifting method[7] and epipolar imaging method[15],[26].

**Micro-phase shifting method.** Micro-phase shifting method was performed to the scenes in Section 6.4, and the 3D reconstruction results are shown in Fig. 9. For the three inter-reflection scenes, the projected patterns for micro-phase shifting were chosen as the 16-15 pattern set, which contains patterns for a frequency-band around 16 pixels and 15 frequencies. For the three subsurface scattering scenes, the projected patterns for micro-phase shifting were chosen as the 64-15 pattern set, which contains patterns for a frequency-band around 64 pixels, and 15 frequencies. The results reveal that the micro-phase shifting method failed to reconstruct complete 3D data for the following scenes: the scenes of a gypsum bear with a mirror and the aluminum alloy workpiece. High-frequency inter-reflections induce micro-phase shifting method failure. The micro-phase shifting method reconstructed complete 3D point for other scenes, but the quality of the point clouds was markedly reduced. Ripple errors were observed on these scenes. By contrast, quick pPSI achieved complete and high-quality 3D data for these scenes.

The superior capability of pPSI is attributed to its explicit separation of the influences of global and direct illumination, enabling 3D data reconstruction under remarkably complex global illumination. Moreover, pPSI projects low- and high-frequency patterns, and all the response information is synthesized by Fourier transform. Thus, pattern set selection is no longer necessary. Thus, pPSI can reconstruct interreflections and strong subsurface scattering simultaneously (with the same pattern set to overcome interreflections and subsurface scattering).

**Epipolar imaging method.** Epipolar imaging was originally conducted by an ad-hoc hardware containing a pair of digital micro-mirror (DMD) devices for projection and masking[15]. These devices achieved various probing ma-trix by simultaneously controlling patterns on both DMDs at a high frequency. One of their goals was to form the epipolar-only imaging to achieve 3D reconstruction under global illumination. This goal was based on the non-epipolar dominance assumption, which states that epipolar global illumination is weak. However, this assumption can be easily violated, as in the experiment conducted in Fig. 10. Fig. 10 (a) shows that the V-groove is in a vertical configuration mode. Interreflections mostly occur between the right and the left planes, and the configuration of the proposed 3D system [Fig. 3 (d)] facilitates closeness of the epipolar line to a horizontal line. Thus, epipolar global illumination is dominated in this setting. The epipolar dominated region is shown in Fig. 10 (b). Rather than using the specially designed hardware in [15], the epipolar imaging method introduced in [26], which adopted the same imaging setup with this paper, is utilized. Paper [26] captured one image per pattern and then performed masking in the post process, thereby also facilitating 3D reconstruction under global illumination. The 3D reconstruction results by using epipolar imaging are shown in Fig. 10 (c). A large region of missing data due to epipolar global illumination is observed. Quick pPSI was also conducted for the same scene. The imaging parameters is the same as in Section 6.4. Complete and high-quality data can then be obtained. This experiment confirmed that the proposed method is robust to epipolar dominated global illumination.

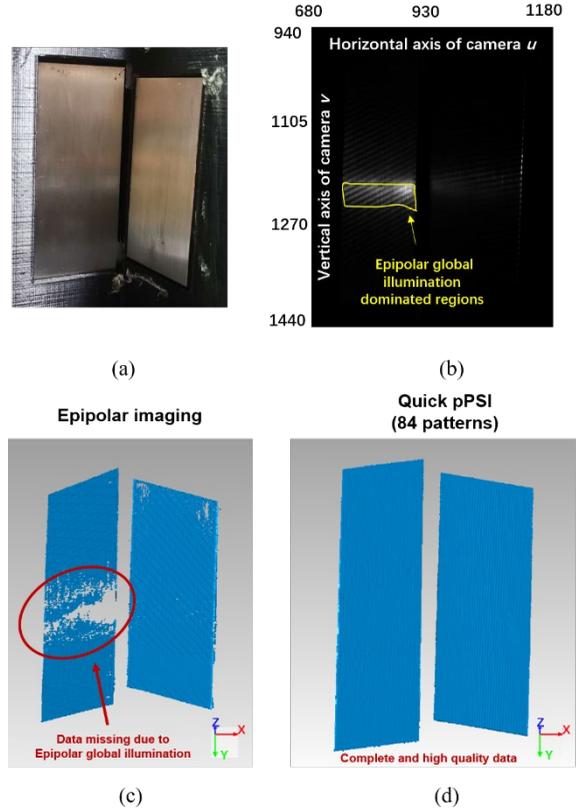

(a)                                    (b)

(c)                                    (d)

Fig. 10. Comparison with epipolar imaging method. (a) Investigated scene. The V-groove is in a vertical configuration mode such that epipolar global illumination is dominated. (b) V-groove from camera perspective, where the epipolar global illumination dominated region was shown by the solid line. (c) 3D reconstruction result by epipolar imaging. The missing data were caused by epipolar global illumination. (d) 3D reconstruction result by quick pPSI, where complete and high-quality points are obtained.

# 7 CONCLUSION

pPSI is introduced in the present paper for efficient and robust correspondence matching in instances dominated by global illumination. The relationship between LTC and projection functions is theoretically demonstrated. The oblique sinusoidal pattern illumination mode is also proposed. The local maximum constraint is introduced to identify the candidate correspondence points by intersecting the region that satisfies the local maximum constraint. The LSE method is presented to further accelerate capture efficiency. Several challenging scenes are measured and compared, and results validate that pPSI achieves efficient and robust 3D shape measurement in the presence of global illumination. Optimizations are also conducted for pPSI. First, the number of projection functions required for pPSI is optimized for several situations. Second, the influence of capture ratio in LSE on the accuracy of the correspondence matching points is investigated. Quick pPSI is the method that combines both of these optimization strategies, which can achieve 3D reconstruction under global illumination with dozens of projected patterns.

## ACKNOWLEDGMENT

This work was supported by Key Research and Development Program of China (2020YFB2010701). The authors wish to thank Yang Xu for providing assistance to conduct the experiment of epipolar imaging, and Lu Wang for providing assistance to build the 3D imaging system.

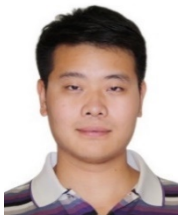

**Yuxi Li** received the BS degree from North China Electric Power University, China, in 2013, MS degree from Tianjin University, China, in 2017, and PhD degree from Beihang University, China, in 2023. His current research interests include computational imaging, 3D computer vision and computer graphics.

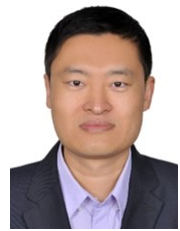

**Hongzhi Jiang** received the PhD degree from Beihang University, China, in 2010. He currently works as an associate professor at the School of Instrumentation and Optoelectronic Engineering, Beihang University, China. His current research interests include optical 3D measurement, computer vision.

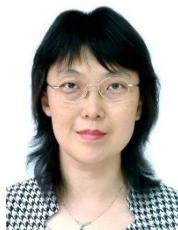

**Huijie Zhao** received her PhD degree from Harbin Institute Technology, China, in 1994. Now she works as a professor at the Institute of Artificial Intelligence, Beihang University, China. She is also the vice-director of the Key Laboratory of Precision Opto-mechatronics Technology, Ministry of Education, China. Her research interests include 3-D measurement techniques and application in Industry, Hyperspectral imaging techniques, Optical Imaging System Modelling, Simulation and evaluation, Hyperspectral data Processing, etc. She has published more than 200 research papers and hold more than 100 National Invention Patents of China. She has supervised more than 30 PhD candidates so far.

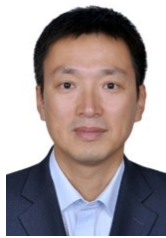

**Xudong Li** received the PhD degree from Beihang University, China, in 2003. He currently works as a professor at the School of Instrumentation and Optoelectronic Engineering, Beihang University, China. His current research interests include optical 3D measurement, computer vision.

# Projective Parallel Single-pixel Imaging: 3D Structured Light Scanning Under Global Illumination

## *** Appendices ***

Yuxi Li*, Hongzhi Jiang*, Huijie Zhao §, and Xudong Li

- - - - - - - - - ◆ - - - - - - - - -

## APPENDIX A

### PROOF OF LOCAL MAXIMUM CONSTRAINT PROPOSITION

Local maximum constraint, which states that the correspondence matching point must be a local maximum on the projection function, provides necessary condition for the location of candidate correspondence matching points when projection functions are captured. A lemma is required to prove the local maximum constraint proposition.

**Lemma 1 Derivative properties of Radon transform.** *The Radon transform of the original function derivative has the form of*

$$\Re_\theta\{\frac{\partial f}{\partial x}\} = \cos\theta\frac{\partial \check{f}(\rho,\theta)}{\partial \rho},$$

$$\Re_\theta\{\frac{\partial f}{\partial y}\} = \sin\theta\frac{\partial \check{f}(\rho,\theta)}{\partial \rho}, \quad (A1)$$

*and the Radon transform of the second derivative of the original function has the form of*

$$\Re_\theta\{\frac{\partial^2 f}{\partial x^2}\} = \cos^2\theta\frac{\partial^2 \check{f}(\rho,\theta)}{\partial \rho^2},$$

$$\Re_\theta\{\frac{\partial^2 f}{\partial y^2}\} = \sin^2\theta\frac{\partial^2 \check{f}(\rho,\theta)}{\partial \rho^2}. \quad (A2)$$

where the Radon transform of a 2D function $f(x)$ is denoted as $\Re_\theta f(x,y) = \check{f}(\rho,\theta)$.

*Proof* can be found in [36].

**Local Maximum Constraint Proposition**. *If the corresponding projection line does not pass through any speckles due to global illumination, then the direct illumination point on the pixel transport image is a local maximum point on the projection functions.*

---

- *: Equal Contribution
- Yuxi Li, Hongzhi Jiang and Xudong Li are with School of Instrumentation and Optoelectronic Engineering, Beihang University (BUAA), Beijing 100191, China.
  E-mail: uniluxli@ qq.com, jhz1862@buaa.edu.cn, xdli@buaa.edu.cn.
- Huijie Zhao is with Institute of Artificial Intelligence, Beihang University (BUAA), Beijing 100191, China. E-mail: hjzhao@buaa.edu.cn.
- § Corresponding author: Huijie Zhao.

*Proof* Suppose $(x_0, y_0)$ is the direct illumination point, which is a local maximum point on the pixel transport image, and the small circular region $\Omega_R$ with radius $R$ centered at $(x_0, y_0)$ is the effective region of the direct illumination (Fig.A1). This effective region is caused by blurring effect of the lens; thus, the value of pixel transport image inside $\Omega_R$ can be regarded as axial symmetric and concave function. Therefore, a radius square $r^2 = (x - x_0)^2 + (y - y_0)^2$ is sufficient to parameterize the function inside $\Omega_R$

$$f(x,y) = f(r^2),$$

$$\frac{\partial f(x,y)}{\partial x} = 2f'_{r^2}x,$$

$$\frac{\partial f(x,y)}{\partial y} = 2f'_{r^2}y, \quad (A3)$$

$$\frac{\partial^2 f(x,y)}{\partial x^2} = 4f''_{r^2}x^2 + 2f'_{r^2},$$

$$\frac{\partial^2 f(x,y)}{\partial y^2} = 4f''_{r^2}y^2 + 2f'_{r^2},$$

where $f(r^2)$ is a concave and monotone decreasing function considering $r^2$ and $f'_{r^2} < 0, f''_{r^2} < 0$, which are the derivative and second derivative of function $f(\bullet)$ considering $r^2$.

Eq. (A3) demonstrates the following conclusions

$$\iint_{\Omega_R}\frac{\partial f(x,y)}{\partial x}\cdot\delta(\rho_0 - x\cos\theta - y\sin\theta)dxdy = 0,$$

$$\iint_{\Omega_R}\frac{\partial f(x,y)}{\partial y}\cdot\delta(\rho_0 - x\cos\theta - y\sin\theta)dxdy = 0, \quad (A4)$$

and

$$\iint_{\Omega_R}\frac{\partial^2 f(x,y)}{\partial x^2}\cdot\delta(\rho_0 - x\cos\theta - y\sin\theta)dxdy < 0,$$

$$\iint_{\Omega_R}\frac{\partial^2 f(x,y)}{\partial y^2}\cdot\delta(\rho_0 - x\cos\theta - y\sin\theta)dxdy < 0, \quad (A5)$$

where $\rho_0 = x_0\cos\theta + y_0\sin\theta$. Eqs. (A4) and Eq. (A5) indicate that the integral is only applied along the line $L(\rho_0, \theta): x\cos\theta + y\sin\theta - \rho_0 = 0$ because the delt function is not equal to zero only along this line.

Eq. (A4) holds because $\partial f/\partial x|_{x=x_0+\delta}=-\partial f/\partial x|_{x=x_0-\delta}$ and $\partial f/\partial y|_{y=y_0+\delta}=-\partial f/\partial y|_{y=y_0-\delta}$ can be concluded from Eq. (A3). Eq. (A5) holds because $\partial^2 f(x,y)/\partial x^2<0$ and $\partial^2 f(x,y)/\partial y^2<0$ can be concluded according to Eq. (A3).

Eqs. (A4) and Eq. (A5) can be understood as the integral of the corresponding partial derivative functions along line $x\cos\theta+y\sin\theta-\rho_0=0$ , where $\rho_0=x_0\cos\theta+y_0\sin\theta$ . It is a line passing through point $(x_0,y_0)$ , and the angle between $x$ axial is $\theta$ . $\rho_0$ is the projection point of $(x_0,y_0)$ .

From the definition of Radon transform, the Radon transform of the derivative of the original function can be written as

$$\Re_\theta \frac{\partial f}{\partial x}=\iint[\frac{\partial f}{\partial x}\cdot\delta(\rho-x\cos\theta-y\sin\theta)]dxdy$$

$$\Re_\theta \frac{\partial f}{\partial y}=\iint[\frac{\partial f}{\partial y}\cdot\delta(\rho-x\cos\theta-y\sin\theta)]dxdy. \tag{A6}$$

The projection line does not pass through any speckles caused by global illumination for a certain $\theta$ ; thus, the values of function $f(x,y)$ (also function $\partial f(x,y)/\partial x$ ) on line $L(\rho_0,\theta):x\cos\theta+y\sin\theta-\rho_0=0$ are zero, except for the partial line in region $\Omega_R$ . For the integral inside $\Omega_R$ , the integral along a fixed line $\rho_0=x\cos\theta+y\sin\theta$ in Eq. (A6) is equal to zero, according to Eq. (A4). Eq. (A1) then reveals the following:

$$\Re_\theta \frac{\partial f}{\partial x}|_{(x,y)\in L(\rho_0,\theta)}=\cos\theta\frac{\partial \breve{f}(\rho,\theta)}{\partial \rho}|_{\rho=x_0\cos\theta+y_0\sin\theta}=0,$$

$$\Re_\theta \frac{\partial f}{\partial y}|_{(x,y)\in L(\rho_0,\theta)}=\sin\theta\frac{\partial \breve{f}(\rho,\theta)}{\partial \rho}|_{\rho=x_0\cos\theta+y_0\sin\theta}=0. \tag{A7}$$

For any value of $\theta$ , $\sin\theta$ and $\cos\theta$ are not simultaneously equal to zero. Thus, the following conclusion is presented:

$$\frac{\partial \breve{f}(\rho,\theta)}{\partial \rho}|_{\rho=x_0\cos\theta+y_0\sin\theta}=0 . \tag{A8}$$

From the definition of Radon transform, the second derivative of the original function can be written as

$$\Re_\theta \frac{\partial^2 f}{\partial x^2}=\iint[\frac{\partial^2 f}{\partial x^2}\cdot\delta(\rho-x\cos\theta-y\sin\theta)]dxdy$$

$$\Re_\theta \frac{\partial^2 f}{\partial y^2}=\iint[\frac{\partial^2 f}{\partial y^2}\cdot\delta(\rho-x\cos\theta-y\sin\theta)]dxdy. \tag{A9}$$

Based on a similar derivation between Eqs. (A6) and (A7) and Eq. (A5), Eq. (A9) can be written as

$$\Re_\theta \frac{\partial^2 f}{\partial x^2}|_{(x,y)\in L(\rho_0,\theta)}=\cos^2\theta\frac{\partial^2 \breve{f}(\rho,\theta)}{\partial \rho^2}|_{\rho=x_0\cos\theta+y_0\sin\theta}$$

$$=\iint_{\Omega_R}[\frac{\partial^2 f}{\partial x^2}\cdot\delta(\rho_0-x\cos\theta-y\sin\theta)]dxdy<0$$

$$\Re_\theta \frac{\partial^2 f}{\partial y^2}|_{(x,y)\in L(\rho_0,\theta)}=\sin^2\theta\frac{\partial^2 \breve{f}(\rho,\theta)}{\partial \rho^2}|_{\rho=x_0\cos\theta+y_0\sin\theta}$$

$$=\iint_{\Omega_R}\frac{\partial^2 f(x,y)}{\partial y^2}\cdot\delta(\rho_0-x\cos\theta-y\sin\theta)dxdy<0. \tag{A10}$$

$\sin^2\theta\geq 0$ and $\cos^2\theta\geq 0$ ; thus,

$$\frac{\partial^2 \breve{f}(\rho,\theta)}{\partial \rho^2}|_{\rho=x_0\cos\theta+y_0\sin\theta}<0 . \tag{A11}$$

Therefore, Eqs. (A8) and (A11) reveal that the projection point $\rho_0$ is a local maximum on the projection function $\breve{f}(\rho,\theta)$ . □

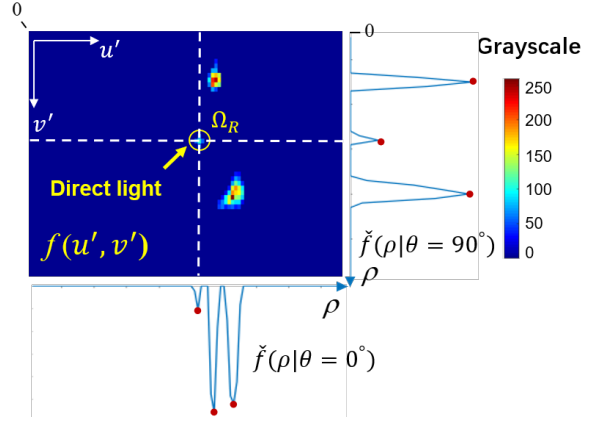

Fig. A1: Direct light and its nearby region.

# APPENDIX B

## PERFECT RECONSTRUCTION OF PROJECTION FUNCTIONS BY LOCAL SLICE EXTENSION METHOD

In the main text, the projection functions are captured by the proposed local slice extension (LSE) method, which is implemented by a "coarse to fine" localization procedure. LSE will be introduced in this section from a theoretical aspect, and the capability of the LSE method to reconstruct the projection function perfectly is proven. First, Lemma 2 is provided. Lemma 2 states that the reconstructed projection function by inverse discrete Fourier transform (IDFT) corresponds to a periodic extension version of the projection function when the patterns generated by Eq. (9) are projected.

**Lemma 2** *Assume $f_\theta(\rho;u,v)$ is the projection function with direction $\theta$ for camera pixel $(u,v)$ . By projecting patterns in the form of Eq. (9), the reconstructed function of camera pixel $(u,v)$ by IDFT becomes a periodic extension version of the original projection function*

$$\tilde{f}_\theta(\rho;u,v)=2b\cdot\sum_{r_1=-\infty}^{+\infty}f(\rho_r-r_1 M_\theta;u,v), \tag{A12}$$

*where $\rho_r$ is a pixel on the reconstructed function, $M_\theta$ is given in Eq. (9), which is the size of the maximum of $\theta$ projected reception field for each camera pixel, and $r_1$ is integer.*

**Proof** Similar to Eqs. (7) and (8), when each sample in the frequency domain is obtained by using the patterns generated by Eq. (9), the reconstructed function by applying IDFT on the captured intensity is calculated as

$$\tilde{f}_\theta^r(\rho_r; u, v) = IDFT\{\frac{Sb}{2} \cdot \sum_{v'=0}^{N-1}\sum_{u'=0}^{M-1} \cdot h(u', v'; u, v)$$

$$\cdot \exp[-\frac{2\pi k}{M_\theta}(u'\cos\theta + v'\sin\theta)]\}$$

$$= \frac{Sb}{2} \cdot \sum_{r_1=-\infty}^{+\infty}\sum_{v'=0}^{N-1}\sum_{u'=0}^{M-1} \cdot h(u', v'; u, v)$$  (A13)

$$\cdot \delta(\rho - u'\cos\theta - v'\sin\theta - r_1 M_\theta)$$

$$= \frac{Sb}{2} \cdot \sum_{r_1=-\infty}^{+\infty} f_\theta^{Radon}(\rho - r_1 M_\theta; u, v)$$

$$= \sum_{r_1=-\infty}^{+\infty} f_\theta^P(\rho - r_1 M_\theta; u, v),$$

where $f_\theta^{Radon}(\rho; u, v)$ is the discrete Radon transform of LTC along direction $\theta$, as defined in Eq. (2).  □

The periodic extension version of projection function is shown in Fig. A2. The LSE reconstruction theorem can be proven using Lemma 2.

**LRE Reconstruction Theorem**. *If the size of the maximum of $\theta$ projected reception field $M_\theta$ covers the non-zero regions of $f_\theta(\rho; u, v)$, then projection function can be perfectly reconstructed by adopting the LSE method. That is, the projection function obtained by the LSE method implemented by the "coarse to fine" procedure is exactly equal to the projection function captured and reconstructed in accordance with Eqs. (4)–(8).*

*Proof* If the size of the maximum of $\theta$ projected reception field $M_\theta$ covers the non-zero regions of $f_\theta(\rho; u, v)$, then the reconstructed projection function from Lemma 2 can be regarded as a periodic extension version of the original projection function, with step size of $M_\theta$. Aliasing does not occur in this situation (Fig. A2), and all information of $f_\theta(\rho; u, v)$ is preserved in $\tilde{f}_\theta^r(\rho; u, v)$. Thus, projection function, which is equal to projection function captured and reconstructed in accordance with Eqs. (4)–(8), can be exactly reconstructed by adopting the LSE method.

Provided that the nonzero region $C_\theta(\rho; u, v)$ of $f_\theta(\rho; u, v)$ is known, the projection function $f_\theta^r(\rho; u, v)$ can be exactly obtained by taking the values inside the visible region $C_\theta(\rho; u, v)$ of $\tilde{f}_\theta^r(\rho; u, v)$ and setting zeros outside $C_\theta(\rho; u, v)$ of $\tilde{f}_\theta^r(\rho; u, v)$ as given by Eq. (12).  □

## APPENDIX C

## DERIVATION OF THE COARSE LOCALIZATION ACCURACY CONSIDERING FREQUENCY NUMBER

Suppose the projection function $f(\rho)$ has a length of $L$. The truncation process of the high frequencies of the projection function $f(\rho)$ can be described as applying a window function $P(k)$ on $F(k)$, which is the discrete Fourier transform (DFT) of $f(\rho)$

$$F^C(k) = P(k)F(k),$$  (A14)

where $F^C(k)$ is the captured frequencies for coarse localization, and $P(k)$ has the following form

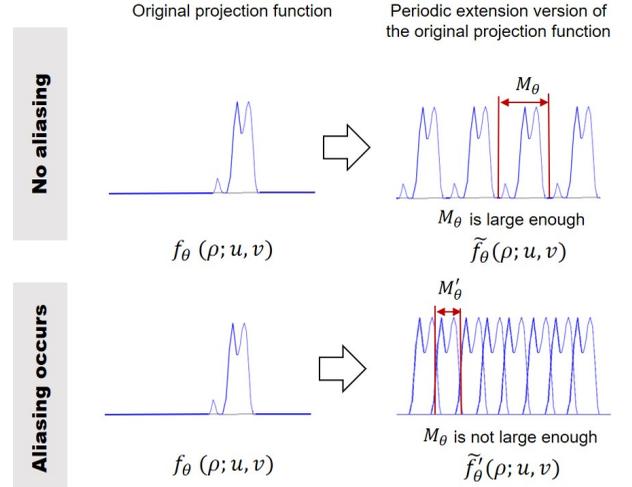

Fig. A2: Perfect reconstruction property of the LSE method. Periodic extension version of projection function is shown. In the first row, the $M_\theta$ is sufficiently large such that aliasing does not occur. While in the second row, $M_\theta$ is insufficiently large, thus allowing the occurrence of aliasing.

$$P(k) = \begin{cases} 1 & 0 \le k \le K-1 \text{ or } L\text{-}K\text{ -}1 \le k \le L-1 \\ 0 & \text{otherwise} \end{cases},$$  (A15)

where $K$ is the captured number of low frequencies. Conjugate symmetry property is considered, resulting in the two banded nonzero values in Eq. (A15).

From circular convolution property, Eq. (A14) is equivalent to applying circular convolution in the spatial domain

$$F^C(k) = P(k)F(k) \overset{DFT}{\leftrightarrow} f^C(\rho) = p(\rho) \otimes_L f(\rho),$$  (A16)

where $\otimes_L$ denotes L-point circular convolution. $\overset{DFT}{\leftrightarrow}$ denotes that the two functions are a DFT pair. If $p(\rho)$ is a periodic impulse train with period L, then $f^C(\rho)$ is exactly equal to $f(\rho)$. This condition corresponds to capturing the overall frequencies $k = 0, \cdots, L-1$.

When truncation is applied, the reconstructed coarse projection function $f^C(\rho)$ is the L-point circular convolution of $p(\rho)$ and $f(\rho)$, as shown in Figs. A3 (a)–(c). The approximated size of the projected reception field $M_s$ is determined by the width of the main lobe of $p(\rho)$, which is depicted in Fig. A3 (b). The length of $M_s$ can be approximated by adding the actual size of the projected reception field and the size of the main lobe.

$p(\rho)$ can be obtained by applying IDFT to $P(k)$, as expressed by

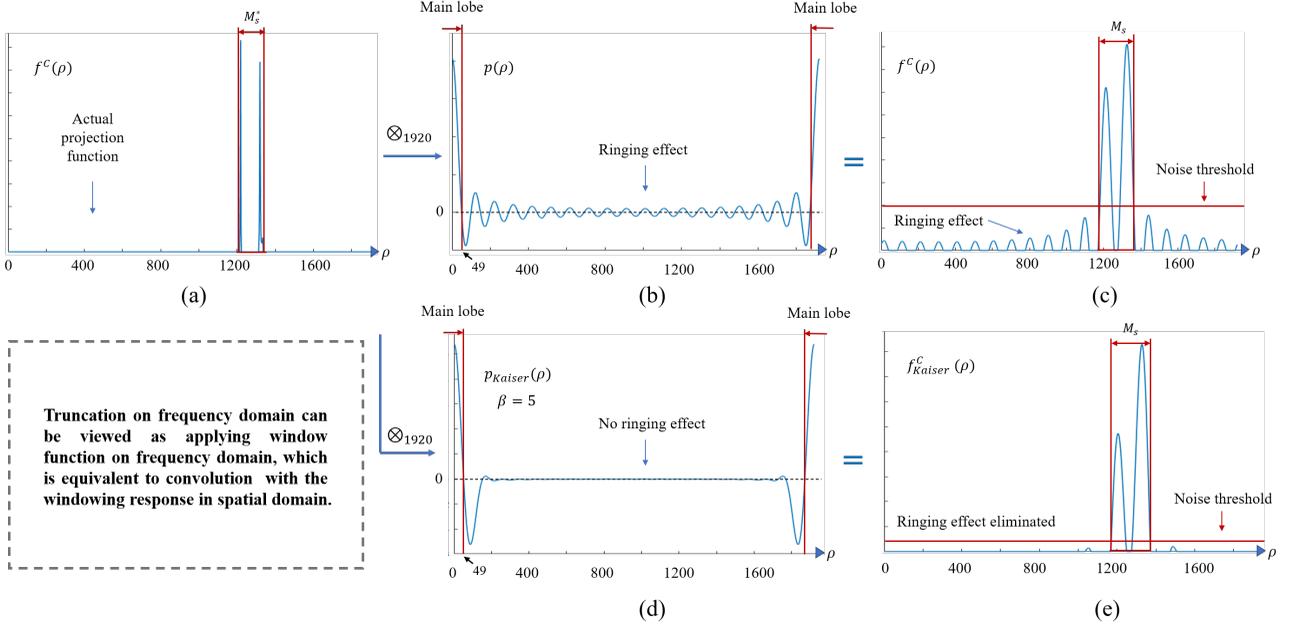

Fig. A3: Truncation in frequency domain. (a) Actual projection function. The projected reception field is indicated by the red lines. (b) Convolution kernel in spatial domain when direct truncation is applied in frequency domain. The size of the main lobe is important for localization accuracy. Ringing effect is observed if high frequency is directly removed. (c) The reconstructed coarse projection function when direct truncation is applied. The location of the projected reception field can be determined by setting the region where the values are larger than a noise threshold. (d) Convolution kernel in spatial domain when Kaiser window, with the shape parameter set as 5, is applied in frequency domain. Ringing effect is absent. (e) The reconstructed coarse projection function when Kaiser window is applied in frequency domain.

$$p(\rho) = \frac{1}{L} \sum_{k=0}^{K-1} \exp(j\frac{2\pi}{L}k\rho)$$
$$+ \frac{1}{L} \sum_{k=L-K-1}^{L-1} \exp(j\frac{2\pi}{L}k\rho)$$
$$= \frac{1}{L} \sum_{k=-(K-1)}^{K-1} \exp(j\frac{2\pi}{L}k\rho) \qquad (A17)$$
$$= \frac{1}{L} \frac{\sin[\frac{2\pi}{L}(K-\frac{1}{2})\rho]}{\sin(\frac{\pi}{L}\rho)}.$$

The size of the main lobe can be determined by setting Eq. (A17) equals to zero. The first zero-crossing point $\rho = L/(2K-1)$ is half of the size of the main lobe. For a fixed $L$, the main lobe area is highly concentrated with the capture of additional low-frequency information. Thus, the coarse localization process is highly accurate.

Overall, for a projection function $f(\rho)$ with a length of $L$, the approximated size of the projected reception field $M_s$ is $2L/(2K-1)$ wider than the actual size when $K$ low frequencies are captured.

A tradeoff exists between the number of captured low frequencies and the localization accuracy. To solve this problem, suppose $M_s^*$ is the actual size of the projected reception field, and

$$R = \frac{2L}{(2K-1)M_s^*}, \qquad (A18)$$

is the localization uncertain ratio. $R$=1 in the experiments, which indicates that the approximated size of the projected reception field $M_s$ is only one time wider than the actual size, is enough. $L$ takes a value of 1920, and $M_s^*$ is assumed as 150 pixels. Putting these values into Eq. (A18), $K \approx 13$ is obtained. Results showed that the number of low-frequency samples of 10 is sufficient in most situations.

However, removing high frequency information directly results in severely ringing effect, as shown in Figs. A3 (b)–(c). Alternatively, Kaiser window[37], which has a shape parameter $\beta$ of 5, is applied to the tested low frequency samples. The ringing effect can be largely eliminated, as shown in Figs. A3 (d)–(e). Kaiser window is originally applied on spatial/time domains. However, in the context of pPSI, Kaiser window should be applied on frequency domain.

## APPENDIX D

## RANDOM SAMPLE CONSENSUS (RANSAC) BASED INTERSECTION METHOD WITH FOUR PROJECTIONS

The use of the RANSAC (Random Sample Consensus)-based intersection method with four projections is investigated in this section to solve direct correspondence points. Under the condition that virtual correspondence points can be excluded, projection functions in at least three directions are required. However, considering the possible phenomenon of mixed peaks, at least four projection functions are needed to recognize the wrong direction and obtain the correct result. A method for solving interreflection correspondence points by traversal consistency matching, which is based on RANSAC, is proposed in this section. The idea of using the minimum sample set and checking the consistency is used to eliminate the peaks that may be affected by the mixed peak phenomenon and 3D reconstruction quality.

The specific process of solving method for inter-reflections correspondence points by traversal consistency matching is as follows.

(1) According to the projective parallel single-pixel imaging method based on projection reconstruction described in Section 4, the scene to be measured is captured, the projection fringe patterns in four directions are used for illuminating, and the four directions can adopt the distribution form as $\theta = 0°, 45°, 90°, 135°$;

(2) For the certain pixel $(u, v)$ in the camera, the projection function $f_\theta(\rho; u, v)$ in four directions is calculated by the method presented in Section 4.3. Meanwhile, we define a linked list $D_R$ that maintains a combination of existing peak values. Each node in the link stores a peak combination tuple $i, j, k, w$. The linked list also supports the function of adding new nodes at any time and determining whether a node exists in the linked list, given a node with tuple $i, j, k, w$. The letters $i, j, k, w$ in this tuple denote the $i, j, k, w$-th local maximum value of the projection function when the projection direction $\theta$ is 0°, 45°, 90°, and 135°, respectively, and the intersection can form a correspondence point. When one or more of the letters $i, j, k, w$ is −1, then the projection function is not included in the calculation of the correspondence points to exclude the effect caused by the phenomenon of mixed peaks. The pixels $(u, v)$ in this step and subsequent steps refer to the coordinates after the removal of camera distortion.

(3) Traverse the entire set of two pairwise combinations of projection functions $f_\theta(\rho; u, v)$ in four directions. $\left\{ f_{\theta_1}(\rho; u, v), f_{\theta_2}(\rho; u, v) \right\}$ is regarded as the set after combining certain two projection functions. The program terminates when all projection function combinations have been completed. If all projection function combinations are not traversed, then the following steps are performed.

(4) For two projection functions $\left\{ f_{\theta_1}(\rho; u, v), f_{\theta_2}(\rho; u, v) \right\}$ under a certain combination, all the local maximum combinations in the two projection functions are traversed by a two-fold loop, and $(u'_z, v'_z)$ is the combination of the i-th local maximum point in the first projection function $f_{\theta_1}(\rho; u, v)$ and the j-th local maximum point in the second projection function $f_{\theta_2}(\rho; u, v)$. The matching coordinates $(u'_z, v'_z)$ are calculated in accordance with Eq. (3), and the following steps are performed. If all the local maximum combination has been traversed, then return to the step (3) to implement the next pair of projection function combination.

(5) The epipolar constraint for matching coordinates $(u'_z, v'_z)$ is verified. If the epipolar constraint remains unsatisfied, then return to step (4) and calculate the next set of local maximal combinations. If the epipolar constraint is passed, then the following steps are performed.

(6) According to the following formula,
$$\rho_i = \cos\theta_i \cdot u'_z + \sin\theta \cdot v'_z \tag{A19}$$
where the matching coordinates $(u'_z, v'_z)$ are projected into the two other projection functions to obtain two projection points $\rho_1$ and $\rho_2$. If $\rho_1$ or $\rho_2$ is within the preset fault-tolerant range $\varepsilon_R$ from a certain local maximum coordinate in its corresponding projection function, then the coordinate of the maximum value is retained; if the projection point is not within this preset range, then the indicator of the corresponding direction should be recorded as −1. The preset range $\varepsilon_R$ is 0.5 pixels. If no local maximum values exist in the neighbor of both projection points, then the formed node contains two −1. Step (4) is then performed to calculate the next set of local maximum value combination.

(7) For the node $(i, j, k, w)$ formed in step (6), whether the node $(i, j, k, w)$ already exists is verified. If the node does not exist, then a new location is created and inserted; if the node already exists, then step (4) is subsequently performed to calculate the next set of local maximum value combination.

Algorithm 1 is used as reference for the pseudocode of the algorithm.

# APPENDIX E

## ALGORITHM THAT ADOPTS CONTINUITY CONSTRAINT TO ELIMINATE THE VIRTUALLY MATCHED 3D POINTS

This section describes the recognition method of virtually matched 3D points based on the continuity constraint. As described in Section 5.2, multiple peaks exist in the projection function when interreflections are present in the measurement scene. However, when only the single projection strategy is used, each peak in the projection function is solved with the epipolar constraint to obtain a candidate correspondence point. If the 3D data are solved for all the candidate correspondence points according to the triangulation principle, then many erroneous stray points will emerge, as shown in Fig. 4(b). The 3D data obtained by the virtual matching correspondences are called virtually matched 3D points. Additional constraints must be introduced to determine the virtually matched 3D points. This section studies the recognition and determination of whether a 3D point is a virtually matched 3D point. Thus, the concept of continuity constraint is introduced.

Continuity constraint means that the point cloud of a reasonable measured object should contain sufficient 3D points, and the distance between the 3D points should be remarkably close. Even if the object has several different components that are far apart, each component should have a sufficient number of 3D points, and the 3D points within each component should be sufficiently close to each other. The following definition is required to describe the continuity constraint formally.

<div align="center">

ALGORITHM 1

RANSAC BASED INTERSECTION METHOD WITH FOUR PROJECTIONS

</div>

---

**Solving Correspondence Points by RANSAC based Intersection method with Four Projections**

**Input:** Projection functions $f_{\theta_1}(\rho; u, v)$, $f_{\theta_2}(\rho; u, v)$, $f_{\theta_3}(\rho; u, v)$ and $f_{\theta_4}(\rho; u, v)$ for four directions of a camera pixel $(u, v)$, where the number of local maximum value of each projection function is I, J, K and W, respectively; a pre-set tolerance range $\varepsilon_R$;

**Output:** Linked list $D_R$ of pixel peak combinations, where each node is a tuple of four numbers of the form $(i, j, k, w)$, the meaning of the tuple is: where the letters $i,j,k,w$ denotes the combination of the $i,j,k,w$-th local maximum value of the projection function when the projection direction θ is 0°,45°,90°,135°, respectively, and after the combination of the intersection, a correspondence point is formed.

---

1: Initialize the empty linked list $D_R$;

**Begin traversing all cases of pairwise combinations of projection functions:**

2:　for $f1$ is sequentially cyclically assigned to one of the four projection functions $f_{\theta_1}(\rho; u, v)$, $f_{\theta_2}(\rho; u, v)$, $f_{\theta_3}(\rho; u, v)$ and $f_{\theta_4}(\rho; u, v)$

3:　　for $f2$ is cyclically assigned in order from the latter projection function of $f1$

　　　**The process of traversing the respective local maxima of the combined projection functions begins:**

4:　　　for $i$ is cyclically assigned to the local maximum value in the projection function corresponding to $f1$

5:　　　　for $j$ is cyclically assigned to the local maximum value in the projection function corresponding to $f1$

6:　　　　　if the correspondence point $(u_z', v_z')$ obtained by the intersection after the back projection of the local maximum corresponding to $(i, j)$ satisfies the epipolar constraint

7:　　　　　　Calculate the coordinates $\rho_1$ and $\rho_2$ after projecting $(u_z', v_z')$ to the other two projection functions $f3$ and $f4$, respectively, according to Eq. (A19)

8　　　　　　　if the distance between $\rho_1$ and a certain local maximum value of the projection function $f3$ is less than $\varepsilon_R$

9:　　　　　　　　Record the position $k$ of this local maximum

10:　　　　　　　else

11:　　　　　　　　The position $k$ of this local maximum is noted as -1

12:　　　　　　　end if

13:　　　　　　if the distance between $\rho_2$ and a certain local maximum value of the projection function $f4$ is less than $\varepsilon_R$

14:　　　　　　　Record the position $w$ of this local maximum value

15:　　　　　　else

16:　　　　　　　The position $w$ of this local maximum is noted as -1

17:　　　　　　end if

18:　　　　　　if $w$ and $k$ are not both -1

19:　　　　　　　According to the projection function from which $i, j, k, w$ is obtained, the tuples are placed to form the $remap(i,j,k,w)$, where the $remap$ function is reordered as the given order according to the origin of the four local maximum values

20:　　　　　　　if $remap(i,j,k,w)$ does not exist in $D_R$

21:　　　　　　　　put it into $D_R$

22:　　　　　　　end if

23:　　　　　　end if

24:　　　　　end if

25:　　　　end for

26:　　　end for

27:　　end for

28:　end for

---

**Definition A.1** $r_{th}$ Adjacent Point.

Set $\boldsymbol{P}$ is a set of 3D points, and $p_i \in \boldsymbol{P}$ is a 3D coordinate point. If $\exists q_i \in \boldsymbol{P}$ satisfies the following formula:

$$dist(p_i, q_i) \le r_{th}, \tag{A20}$$

then $q_i$ is called the $r_{th}$ adjacent point of $p_i$, and $dist(p_i, q_i)$ represents the distance between $p_i$ and $q_i$.

**Definition A.2** $N_{th} - r_{th}$ Split Domain.

For a set of 3D points $\boldsymbol{P}$, if the following conditions are both satisfied:

（1） $\forall p_i \in \boldsymbol{P}$, $\exists q_i \in \boldsymbol{P}$ is the $r_{th}$ adjacent point of $p_i$;

（2） The number of 3D points contained in $\boldsymbol{P}$ larger than $N_{th}$;

The 3D point set $\boldsymbol{P}$ then forms a $N_{th} - r_{th}$ split domain.

According to the two above definitions, the continuity constraint can be formally expressed as follows.

**Definition A.3** Continuity Constraints.

Given the parameter $r_{th}$ of the nearest adjacent point and the number of points in the split domain $N_{th}$, for a certain 3D point $p_i$ and a set of points, the sufficient and necessary condition for satisfying the continuity constraint is that $p_i$ is an element in a certain $N_{th} - r_{th}$ split domain.

**Determination Method of Virtually Matched 3D Points.** The method for recognizing the virtually matched 3D points based on the continuity constraint is established on the basis of the continuity constraint to determine whether a certain 3D point $p_i$ is a virtually matched 3D point. The specific method of determination can be described as follows: if $p_i$ satisfies the continuity constraint, then $p_i$ is not a virtually matched 3D point and should be retained; otherwise it should be eliminated.

Set $\boldsymbol{P}$ is the point cloud data obtained by triangulation principle for the candidate correspondence points determined by the unidirectional projection function and the epipolar constraint, and the specific implementation process of searching and retaining the 3D coordinate points satisfying the continuity constraint is as follows:

(1) Initialization stage: A Kd tree is constructed for $\boldsymbol{P}$ to facilitate the subsequent search of the algorithm for adjacent points within the radius range of the specified point. An empty point cloud list $\boldsymbol{R}$ and $\boldsymbol{C}$, which can store 3D coordinate points, is created. Meanwhile, $\boldsymbol{R}$ is the point cloud data after passing the continuity constraint. A queue $\boldsymbol{Q}$ is initialized, and all 3D points are marked as "unprocessed";

(2) For each point $p_i \in \boldsymbol{P}$, if $p_i$ is in the "unprocessed" state, then put $p_i$ in $\boldsymbol{Q}$ and mark $p_i$ as "processed" and perform the following cyclic process.

(i) The first element $q_i$ in the queue $\boldsymbol{Q}$ is checked, all the $r_{th}$ adjacent points are searched and denote as $\boldsymbol{Q}_i^k$, and $q_i$ is moved to $\boldsymbol{C}$;

(ii) For each $q_i^k \in \boldsymbol{Q}_i^k$, only the points marked as the "unprocessed" state are put in the $\boldsymbol{Q}$ while marking the points as "processed";

(iii) When no elements exist in $\boldsymbol{Q}$, the number of elements in the list $\boldsymbol{C}$ is checked. If the number is less than the pre-set value $N_{th}$, then all the elements in $\boldsymbol{C}$ are virtually matched 3D points, which should be eliminated. Meanwhile, if the number of elements is more than the pre-set value $N_{th}$, then all the elements in $\boldsymbol{C}$ are not virtu-ally matched 3D points, which should be retained, and the element in $\boldsymbol{C}$ should be stored in $\boldsymbol{R}$.

The results before and after the recognition and elimination of virtually matched 3D points by using the above methods are shown in Fig. 4 (main text). The parameter setting of the continuity constraint in the current experiments are as follows: $r_{th} = 0.2$ (mm) and $N_{th} = 20,000$ (Points).

## APPENDIX F

## CALCULATION OF THE NUMBER OF PATTERNS

This section introduces the calculation of the number of patterns required by pPSI. For a slice with certain direction, suppose that $N_c$ is the number of frequencies for coarse location, and $N_f$ is the size of the projection reception field in the fine location step. $\eta$ is the capture ratio in fine localization step. $S$ is the step number. Then, taking account of the conjugate symmetric property of Fourier spectrum of real-valued signal, the required number of patterns $N_t$ is calculated by

$$N_t = \begin{cases} SN_c + \eta(\dfrac{S}{2}N_f) - S & S \text{ is even} \\[2ex] SN_c + \eta\dfrac{S}{2}(N_f+1) - S & S \text{ is odd}, N_f \text{ is odd} \\[2ex] SN_c + \eta S(\dfrac{N_f}{2}+1) - S & S \text{ is odd}, N_f \text{ is even.} \end{cases} \tag{A21}$$